\documentclass[a4paper,11pt]{article}
\usepackage{heppub}

\usepackage{stmaryrd}
\usepackage{comment}
\usepackage{color}
\usepackage{amsmath}
\usepackage{amssymb}
\usepackage{amsthm}
\usepackage{mathrsfs}
\usepackage{graphicx}
\usepackage{marvosym}
\usepackage{amsmath,bm}
\usepackage{array}
\usepackage{latexsym}
\usepackage{enumerate}
\usepackage{slashed}
\usepackage{hyperref}
\usepackage{color}
\usepackage{setspace}
\usepackage[hang,small,bf]{caption}
\usepackage[subrefformat=parens]{subcaption}
\usepackage{bbold}
\allowdisplaybreaks

\newcommand*\diff{\mathop{}\!\mathrm{d}}

\newcommand{\nn}{\nonumber}

\newcommand{\be}{\begin{eqnarray}}
	\newcommand{\ee}{\end{eqnarray}}
\newcommand{\ma}{\mathrm}
\newcommand{\ml}{\mathcal}
\newcommand{\bs}{\boldsymbol}
\newcommand{\Tr}{\mathrm{Tr}}
\DeclareMathOperator{\sign}{sign}

\usepackage[normalem]{ulem} 

\title{Quarkonium Polarization in Medium from Open Quantum Systems and Chromomagnetic Correlators}
\date{\today}

\author[a]{Di-Lun Yang}
\author[b]{and Xiaojun Yao}
\affiliation[a]{Institute of Physics, Academia Sinica, Taipei, 11529, Taiwan}
\affiliation[b]{InQubator for Quantum Simulation, Department of Physics, University of Washington, Seattle,WA 98195, USA}
\emailAdd{dlyang@gate.sinica.edu.tw}
\emailAdd{xjyao@uw.edu}
\abstract{We study the spin-dependent in-medium dynamics of quarkonia by using the potential nonrelativistic QCD (pNRQCD) and the open quantum system framework. We consider the pNRQCD Lagrangian valid up to the order $\frac{r}{M^0}=r$ and $\frac{r^0}{M}=\frac{1}{M}$ in the double power counting. By considering the Markovian condition and applying the Wigner transformation upon the diagonal spin components of the quarkonium density matrix with the semiclassical expansion, we systematically derive the Boltzmann transport equation for quarkonia with polarization dependence in the quantum optical limit. Unlike the spin-independent collision terms governed by certain chromoelectric field correlators, new gauge invariant correlators of chromomagnetic fields determine the recombination and dissociation terms with polarization dependence at the order we are working. We also derive a Lindblad equation describing the in-medium transitions between spin-singlet and spin-triplet heavy quark-antiquark pairs in the quantum Brownian motion limit. The Lindblad equation is governed by new transport coefficients defined in terms of the chromomagnetic field correlators. Our formalism is generic and valid for both weakly coupled and strongly coupled quark gluon plasmas. It can be further applied to study spin alignment of vector quarkonia in heavy ion collisions.}

\preprint{IQuS@UW-21-079}

\begin{document}
	\maketitle
	\flushbottom
\section{Introduction}
For a long time, heavy quarkonia have been considered as useful probes of the quark gluon plasma (QGP) in heavy ion collisions after the early studies of the static plasma screening effect~\cite{Matsui:1986dk,Karsch:1987pv}. More recently, it was realized that the dynamical medium effects such as dissociation~\cite{Laine:2006ns,Beraudo:2007ky,Brambilla:2008cx} and recombination \cite{Thews:2000rj} further modify the quarkonium transport. Over the years, there have been comprehensive studies for the semiclassical transport equations incorporating the aforementioned effects for quarkonia~\cite{Grandchamp:2003uw,Grandchamp:2005yw,Yan:2006ve,Liu:2009nb,Song:2011xi,Song:2011nu,Sharma:2012dy,Nendzig:2014qka,Krouppa:2015yoa,Chen:2017duy,Zhao:2017yan,Du:2017qkv,Aronson:2017ymv,Ferreiro:2018wbd,Yao:2020xzw,Zhao:2024gxt} along with the approach from statistical hadronization models~\cite{Andronic:2007bi,Andronic:2007zu}. Moreover, anisotropic effects of the QGP upon quarkonium transport have been also explored~\cite{Dumitru:2007hy,Dumitru:2009fy,Bhaduri:2020lur}. The modern theoretical framework to study the in-medium dynamics of heavy quark-antiquark ($Q\bar{Q}$) pairs is the open quantum system approach~\cite{Young:2010jq,Borghini:2011ms,Akamatsu:2011se,Akamatsu:2014qsa,Blaizot:2015hya,Katz:2015qja,Brambilla:2016wgg,Brambilla:2017zei,Kajimoto:2017rel,DeBoni:2017ocl,Blaizot:2017ypk,Blaizot:2018oev,Akamatsu:2018xim,Miura:2019ssi,Brambilla:2019tpt,Yao:2018nmy,Yao:2020eqy,Brambilla:2021wkt,Binder:2021otw,Brambilla:2022ynh,Brambilla:2023hkw,Brambilla:2024tqg,Delorme:2024rdo}, in which the system of $Q\bar{Q}$ pairs is regarded as an open system interacting with a thermal environment, i.e., the QGP. Such a formalism can be further combined with the effective field theory (EFT) that integrates out the high-energy degrees of freedom and systematically constructs the interaction between the system and the environment via the separation of different energy scales. The potential nonrelativistic QCD (potential NRQCD or pNRQCD) is one of the useful EFTs for studying quarkonia in a medium~\cite{Brambilla:1999xf,Brambilla:2004jw,Fleming:2005pd}. Notably, by adopting pNRQCD in the open quantum system framework, the semiclassical Boltzmann equation for quarkonia has been derived in the quantum optical limit~\cite{Yao:2018nmy,Yao:2020eqy}. On the other hand, Lindblad equations in the quantum Brownian motion limit have also been derived and widely studied~\cite{Akamatsu:2014qsa,DeBoni:2017ocl,Akamatsu:2018xim,Miura:2019ssi,Brambilla:2019tpt,Brambilla:2021wkt,Brambilla:2022ynh,Brambilla:2023hkw,Brambilla:2024tqg,Delorme:2024rdo}. Classical and quantum numerical methods have been developed to solve the Lindblad equations~\cite{DeJong:2020riy,deJong:2021wsd,Omar:2021kra,Lin:2024eiz}. Recent reviews of the quarkonium transport in the open quantum system approach can be found in Refs.~\cite{Rothkopf:2019ipj,Akamatsu:2020ypb,Yao:2021lus,Sharma:2021vvu}, as well as a summary of latest theoretical developments in Ref.~\cite{Andronic:2024oxz}. However, the polarization dependence of quarkonium transport has not been widely studied especially for nucleus-nucleus collisions except for Refs.~\cite{Cheung:2022nnq,Zhao:2023plc}. Recent studies of $J/\psi$ polarization (alignment) in proton-proton (p-p) or proton-nucleus collisions can be found in Refs.~\cite{Ma:2018qvc,Stebel:2021bbn,Brambilla:2022rjd}.

On the experimental side, a relatively large spin alignment signal of $J/\psi$ with respect to the event plane has been measured in Pb-Pb collisions by the ALICE Collaboration~\cite{ALICE:2022dyy}\footnote{Mostly vanishing spin alignment was measured for $J/\psi$ with respect to the helicity frame and to the Collins-Soper frame at low-transverse momentum~\cite{ALICE:2020iev}.} as opposed to the previous measurements in p-p collisions, where no sizable spin alignment signals for $J/\psi$ were observed at both the LHC~\cite{ALICE:2011gej,ALICE:2018crw} and RHIC~\cite{STAR:2020igu}. On the other hand, the spin alignment for light vector mesons such as $K^{*0}$ and $\phi$ have been also measured~\cite{ALICE:2019aid,STAR:2022fan}. In the coalescence scenario, the spin alignment of vector mesons could stem from the spin correlation of the comprised quark and antiquark forming the vector meson~\cite{Liang:2004xn}. In addition, compared to the relatively weak global spin polarization of $\Lambda$ hyperons mostly originating from the spin polarization of the single strange quark induced by vortical fields~\cite{STAR:2019erd,Becattini:2020ngo}, the unexpectedly large spin alignment signals of vector mesons measured in experiments may infer that the underlying mechanism for spin correlations stems from certain fluctuating sources. Such effects hence do not affect the spin polarization of $\Lambda$ hyperons since the resulting spin polarization also fluctuates and its averaged value vanishes, while the fluctuating spin polarization of the quark and that of the antiquark could be correlated and together yield the nonvanishing spin correlation for spin alignment of the vector meson.
By using the quantum kinetic theory derived from Wigner functions \cite{Chen:2012ca,Hidaka:2016yjf,Hidaka:2017auj,Hattori:2019ahi,Yang:2020hri} (see Ref.~\cite{Hidaka:2022dmn} for a review), the spin alignment from the spin correlation generated by background chromoelectromagnetic fields, such as the turbulent chromoelectromagnetic fields induced by Weibel-type instabilities in the anisotropic QGP~\cite{Mrowczynski:1988dz,Mrowczynski:1993qm,Romatschke:2003ms} or glasma fields in the color-glass-condensate effective theory~\cite{McLerran:1993ni,McLerran:1993ka,McLerran:1994vd,Gelis:2010nm,Lappi:2006fp,Lappi:2006hq}, has been investigated in Refs.~\cite{Muller:2021hpe,Yang:2021fea,Kumar:2022ylt,Kumar:2023ghs}. In particular, the glasma effect may result in large spin alignment of vector mesons at low momentum due to the dominant longitudinal fields that intrinsically generate anisotropic spin correlations. The estimated result is qualitatively consistent with the measurements at the LHC, whereas the magnitude of such an early time effect could be damped by the spin relaxation of quarks in the QGP phase~\cite{Kumar:2023ghs} or be further modified by the turbulent chromoelectromagnetic fields in the anisotropic QGP~\cite{Muller:2021hpe}. See also Ref.~\cite{Hauksson:2023tze} on the jet polarization for the latter case. Overall, the study reveals that the gluonic fluctuations could play a crucial role for spin alignment of vector mesons in heavy ion collisions. See also Refs.~\cite{Sheng:2019kmk,Xia:2020tyd,Goncalves:2021ziy,Sheng:2022wsy,Li:2022neh,Sheng:2022ffb,
Li:2022vmb,Wagner:2022gza,Sheng:2023urn,Dong:2023cng,Fang:2023bbw,Kumar:2023ojl,Sheng:2024kgg} for other possible mechanisms giving rise to spin alignment of vector mesons and Ref.~\cite{Becattini:2024uha} for a recent review.  

Nevertheless, in previous studies, only the light vector mesons are mostly considered and the spin dynamics of quarks and that of antiquarks are treated independently. The spin correlation is retrieved only when the coalescence occurs. Such a simple picture does not work for quarkonium as a bound heavy quark-antiquark pair, since the pair may still remain correlated and cannot be treated independently, especially in the low-temperature QGP phase. To study the spin alignment of $J/\psi$ in heavy ion collisions, it is essential to construct a kinetic theory dynamically tracking the polarization of quarkonium states traveling through the QGP. In this work, we hence generalize the formalism in Refs.~\cite{Yao:2018nmy,Yao:2020eqy} by combining the open quantum system framework and pNRQCD to derive the Boltzmann transport equation with spin dependence for quarkonia in the QGP. A similar approach for the derivation of the kinetic equation for quark spin polarization in a weakly coupled QGP has been pursued in Ref.~\cite{Li:2019qkf}. It is worthwhile to note that the kinetic theory framework with a quasiparticle description is more suitable for quarkonium transport in a medium in comparison with the light flavor. Moreover, we will derive a Lindblad equation to describe the spin dynamics of a heavy quark-antiquark pair at high temperature. The Boltzmann equation and the Lindblad equation enable us to study the time evolution of the polarization for $S$-wave spin-triplet quarkonium states. By adopting the EFT description, we can systematically incorporate nonperturbative effects of a strongly coupled QGP upon the quarkonium polarization dynamics. These effects are encoded in terms of chromomagnetic correlators that are defined in a gauge invariant way. 

The paper is organized as follows: In Sec.~\ref{sect:2}, we will briefly review the open quantum system framework in the quantum optical and the quantum Brownian motion limits and introduce the spin density matrix for vector quarkonium states. Chromomagnetic interactions in pNRQCD that are relevant for quarkonium spin dynamics will be explained in Sec.~\ref{sect:3}, followed by the derivation of the semiclassical spin-dependent Boltzmann equation for quarkonium in Sec.~\ref{sect:Boltzmann_eq} and the derivation of the Lindblad equation in Sec.~\ref{sect:5}. We will summarize and draw conclusions in Sec.~\ref{sect:summary}.

Throughout this paper we use 
the mostly minus signature of the Minkowski metric $\eta_{\mu\nu} = {\rm diag} (1, -1,-1,-1)  $ 
and the completely antisymmetric spatial tensor $\varepsilon_{ijk}$ with $\varepsilon_{123}=1$.
Greek and roman indices are used for spacetime and spatial components, respectively, unless otherwise specified. Bold letters will be used to indicate spatial Euclidean vectors, e.g., ${\bs A}=(A_x, A_y, A_z)$. In the Euclidean case, we do not distinguish the covariant and contravariant vectors. Dot products of two spatial vectors will be written as ${\bs A}\cdot {\bs B} = A_iB_i = -A^iB_i = A_xB_x + A_yB_y + A_z B_z$.

\section{Open quantum systems and spin density matrix}
\label{sect:2}
We first briefly review the open quantum system framework. Details of the framework along our line of thinking can be found in Ref.~\cite{Yao:2021lus}. 
We consider a system described by the Hamiltonian $H_S$ evolving inside an environment described by the Hamiltonian $H_E$. The system and the environment are interacting via $H_I$. The total Hamiltonian is given by
\be
H=H_S+H_E+H_I \,.
\ee
We further assume the interaction Hamiltonian can be written as
\be
\label{eqn:H_I}
H_I = \sum_\alpha O^{(S)}_\alpha \otimes O^{(E)}_\alpha \,,
\ee
where $O^{(S)}_\alpha$ and $O^{(E)}_\alpha$ are Hermitian operators that only act on the system and the environment respectively.
In the interaction picture defined by $H_S+H_E$, i.e., the Hamiltonian free of the system-environment interaction, the time evolution of the total density matrix is given by
\begin{align}
\rho^{(\rm int)}(t) = U(t,t_i) \rho^{(\rm int)}(t_i) U^\dagger(t,t_i) \,, \quad U(t,t_i) = \ml{T} \exp\left( -i\int_{t_i}^t {\rm d}t' H_I^{({\rm int})}(t') \right) \,,
\end{align}
where $\ml{T}$ is the time-ordering operator and 
\begin{align}
\rho^{(\rm int)}(t) = e^{i(H_S+H_E)t} \rho(t) e^{-i(H_S+H_E)t} \,,\quad H_I^{({\rm int})}(t) = e^{i(H_S+H_E)t} H_I e^{-i(H_S+H_E)t} \,.
\end{align}
Tracing out the environment degrees of freedom gives the reduced density matrix of the system $\rho_S^{({\rm int})}(t) = {\rm Tr}_E [\rho^{(\rm int)}(t)] $. In the following, we will assume that the total density matrix is factorized at time $t_i$ and furthermore, the environment density matrix is a thermal (Gibbs) state that is only time dependent through its temperature
\begin{align}
\label{eqn:rho_factorize}
\rho^{(\rm int)}(t_i) = \rho_S^{({\rm int})}(t_i) \otimes \rho_E^{({\rm int})}(t_i) \,,\quad \rho_E^{({\rm int})}(t_i) = \frac{e^{-\beta(t_i) H_E}}{Z_E} = \frac{e^{-\beta(t_i) H_E}}{{\rm Tr}_E(e^{-\beta(t_i) H_E}) } \equiv \rho_T(t_i) \,,
\end{align}
where $\beta=\frac{1}{T}$ is given by the inverse of the environment temperature. The assumption of the density matrix factorization relies on the weak coupling between the system and the environment and the environment being large, the former of which underlies the Markovian approximation, to be used in the derivation of the time evolution equation for $\rho_S(t)$ at the initial time $t_i$, i.e., $\frac{{\rm d}\rho_S(t)}{{\rm d}t}|_{t=t_i}$. If we further assume Eq.~\eqref{eqn:rho_factorize} is valid at any time, i.e., any $t_i$, then the time evolution to be derived will be valid at any time.

Because of the arbitrariness of the initial time $t_i$, we can shift it such that $t_i=0$.
If one expands the time evolution of the density matrix to quadratic order in $H_I$, integrates out the environment degrees of freedom, and neglects the one-point environment correlation function which, if it is not zero, can be made vanish ${\rm Tr}_E(O_\alpha^{(E)}\rho_T) = 0$, one finds\footnote{One needs to make sure all environment operators are excluded from $O_\alpha^{(S)}$.}
\begin{align}
\label{eqn:pre_lindblad}
&\rho_S^{({\rm int})}(t) = \rho_S^{({\rm int})}(0) - \int_0^t {\rm d}t_1 \int_0^t {\rm d}t_2 \frac{{\rm sign}(t_1-t_2)}{2} D_{\alpha\beta}(t_1,t_2) \left[ O^{(S)}_\alpha(t_1) O^{(S)}_\beta(t_2), \rho_S^{({\rm int})}(0) \right] \nn\\
& + \int_0^t {\rm d}t_1 \int_0^t {\rm d}t_2  D_{\alpha\beta}(t_1,t_2) \left( O^{(S)}_\beta(t_2) \rho_S^{({\rm int})}(0) O^{(S)}_\alpha(t_1) - \frac{1}{2}\left\{ O^{(S)}_\alpha(t_1) O^{(S)}_\beta(t_2), \rho_S^{({\rm int})}(0)\right\} \right) \nn\\
& + \ml{O}(H_I^3)\,,
\end{align}
where $\alpha,\beta$ are implicitly summed over. The environment correlation function is defined as
\begin{align}
D_{\alpha\beta}(t_1,t_2) = {\rm Tr}_E\left( O^{(E)}_\alpha(t_1) O^{(E)}_\beta(t_2) \rho_T(0) \right) \,.
\end{align}
Equation~\eqref{eqn:pre_lindblad} is a finite difference equation. In the following, we will discuss under what conditions it turns to a well-defined differential equation. The conditions can be specified by three timescales: the environment correlation time $\tau_E\sim\frac{1}{T}$, the system intrinsic time $\tau_S$ given by the inverse of the system's typical energy gap, and the system relaxation time $\tau_R$ that depends on the coupling strength between the system and the environment. Well-defined time evolution equations for quarkonia can be defined in two limits: The first one is a combination of the quantum optical limit ($\tau_R\gg\tau_E,\tau_R\gg\tau_S$) and the semiclassical limit, in which the time evolution is given by a semiclassical Boltzmann equation. The second one is called the quantum Brownian motion limit ($\tau_R\gg\tau_E,\tau_S\gg\tau_E$) where the time evolution of $\rho_S^{({\rm int})}(t)$ can be written as a Lindblad equation. The common condition of the two limits, i.e., $\tau_R\gg\tau_E$, is the Markovian condition. It is the key to turning the finite difference equation into a well-defined differential time evolution equation, which is realized by coarse graining, i.e., by considering a time step $t$ that satisfies $\tau_R\gg t \gg \tau_E$. Finding such a time step $t$ is only possible under the Markovian condition.

Under one further assumption that the relaxation rate of the system is much larger than the temperature changing rate of the environment, we can treat the density matrix $\rho_T(0)$ in the definition of $D_{\alpha\beta}(t_1,t_2)$ as a time-independent state and then the environment correlation will be translationally invariant $D_{\alpha\beta}(t_1,t_2)=D_{\alpha\beta}(t_1-t_2)$. In practice, this assumption means the quarkonium transition processes happen much faster than the expansion of the QGP and thus the relevant environment state is specified by a local temperature $T(t_i)$ with its gradient corrections negligible.

\subsection{Quantum optical limit}
In this limit, we assume the system relaxation time $\tau_{R}$ is much larger than both the environment correlation time $\tau_{E}$ and the intrinsic timescale of the system $\tau_{S}$, i.e., $\tau_{R}\gg \tau_{E}, \tau_{S}$.  
From Eq.~(\ref{eqn:pre_lindblad}), we may recast the finite difference equation with $t_i=0$ in the Schr\"odinger picture as 
\be 
\rho_S(t) &=& \rho_S(0) -it[H_S,\rho_S(0)] -i\sum_{a,b} \sigma_{ab}(t) [L_{ab}, \rho_S(0) ]  \nn \\
\label{eqn:lindblad}
& + & \sum_{a,b,c,d} \gamma_{ab,cd} (t) \Big( L_{ab} \rho_S(0) L^{\dagger}_{cd} - \frac{1}{2} \{ L^{\dagger}_{cd}L_{ab}, \rho_S(0)\}  \Big)
+ \ml{O}(H_I^3)\,,
\ee
where $L_{ab} \equiv |E_a\rangle \langle E_b|$ and $\{ |E_a\rangle \}$ consists of eigenstates of $H_S$, i.e., $H_S|E_a\rangle = E_a|E_a\rangle$ and forms a complete set of states in the Hilbert space of the system. The explicit expressions of $\sigma_{ab}(t)$ and $\gamma_{ab,cd} (t)$ depend on the detailed interactions dictated by an underlying quantum field theory of the system. With the general interaction term in Eq.~\eqref{eqn:H_I}, each term can be written as
\begin{eqnarray}
&&\sigma_{ab}(t) \equiv \frac{-i}{2}  \sum_{\alpha, \beta} \int_{0}^{t} \diff t_1 \int_{0}^{t} \diff t_2   \sign(t_1-t_2) \langle E_a | O^{(S)}_{\alpha}(t_1) O^{(S)}_{\beta}(t_2) | E_b\rangle D_{\alpha\beta}(t_1, t_2) \,, \nn\\
&&\gamma_{ab,cd} (t) \equiv  \sum_{\alpha, \beta} \int_{0}^{t} \diff t_1 \int_{0}^{t} \diff t_2  \langle E_d | O^{(S)}_{\alpha}(t_1) | E_c \rangle \langle E_a | O^{(S)}_{\beta}(t_2) | E_b \rangle D_{\alpha\beta}(t_1, t_2) \,.
\end{eqnarray}
We will later adopt pNRQCD and manifestly show how the spin degrees of freedom of $Q\bar{Q}$ pairs are involved through the relevant interactions.  

Rigorously speaking, the condition $\tau_R\gg \tau_S$ is not valid for the quantum dynamics of heavy quark pairs, since the energy of the relative motion between an unbound $Q\bar{Q}$ pair is in the continuum. The typical energy gap in the system of unbound pairs can be arbitrarily small but nonzero, which renders the timescale $\tau_S$ arbitrarily long and violates the hierarchy. This means that the $\gamma_{ab,cd}(t)L_{ab}\rho_S(0)L^\dagger_{cd}$ term can receive contributions from the initial state $\langle E_b |\rho_S(0)| E_d \rangle$ where the states $|E_b\rangle$ and $| E_d \rangle$ differ slightly in their relative motion energies. Because of this conceptual issue, a Lindblad equation for quarkonium in the quantum optical limit cannot be easily obtained. However, if a Wigner transformation is applied to the system density matrix, one can recover the hierarchy $\tau_R\gg \tau_S$ under a semiclassical expansion. Details of resolving this issue in the semiclassical limit can be found in Sec.~4.2.2 of Ref.~\cite{Yao:2021lus}.

The Wigner transformation is defined by projecting the density matrix onto certain basis states and then performing a Fourier transform. Given $\rho_S(t)$, we can introduce the polarized distribution function for vector mesons (such as the $S$-wave quarkonium state $J/\psi$ with spin one) through the Wigner transformation\footnote{In principle, one may also consider the off-diagonal components in polarization. However, for simplicity and our phenomenological interest, these terms are ignored in the present study.},
\be
f_{\lambda}({\bs x}, {\bs k}, t) \equiv \int\frac{\diff^3k'}{(2\pi)^3} e^{i {\bs k}'\cdot {\bs x} } \Big\langle  {\bs k}+\frac{{\bs k}'}{2}, \lambda  \Big| \rho_S(t)  \Big|   {\bs k}-\frac{{\bs k}'}{2} , \lambda \Big\rangle \,,
\ee
where we only consider the diagonal component in the polarization (same as spin for $S$-wave quarkonium states) space with $\lambda=+,-,0$ and omit other quantum numbers here for brevity.

One may derive a kinetic equation from Eq.~(\ref{eqn:lindblad}) by employing the hierarchy of timescales and the semiclassical expansion after a Wigner transformation is applied~\cite{Yao:2020eqy}. The term pertinent to $\sigma_{ab}$ is a correction to the system Hamiltonian $H_S$, whose effect can be included when solving the eigenstates $|E_a\rangle$. The $\sigma_{ab}$ term does not lead to a collision term\footnote{The $\sigma_{ab}$ term combined with the $-it[H_{S},\rho_{S}(0)]$ term in Eq.~(\ref{eqn:lindblad}) eventually gives rise to the free-streaming term proportional to the product of the velocity and the spatial derivative upon the distribution function in the Boltzmann equation~\cite{Yao:2018nmy}.}, whereas the $\gamma_{ab,cd}$ terms are responsible for the collision terms in the Boltzmann equation (e.g., the quarkonium recombination and dissociation terms in the spin-independent case). Here we just list the main results, with details given in Sec.~\ref{sect:Boltzmann_eq}. Schematically we define the terms in Eq.~(\ref{eqn:lindblad}) responsible for collisions as
\begin{align}
\label{eqn:Cpm}
\widetilde{\mathcal{C}}^{+}(t) \equiv \sum_{a,b,c,d} \gamma_{ab,cd} (t)L_{ab} \rho_S(0) L^{\dagger}_{cd} \,,\quad
\widetilde{\mathcal{C}}^{-}(t) \equiv \frac{1}{2}\sum_{a,b,c,d} \gamma_{ab,cd} (t) \{ L^{\dagger}_{cd}L_{ab}, \rho_S(0)\} \,.
\end{align}
Under the Markovian condition $\tau_R\gg \tau_E$, we can consider a time step that satisfies $\tau_R\gg t\gg\tau_E$. Using $t\gg\tau_E$, we can show $\widetilde{\mathcal{C}}^{\pm}(t) \propto t$. Then using $\tau_R \gg t$, we can take the limit $t\to0$ in Eq.~\eqref{eqn:lindblad} and obtain a well-defined differential equation. When only the chromomagnetic interactions are considered, the expected kinetic equation takes the form
\begin{align}
\label{eqn:Boltzmann_p}
\frac{\partial}{\partial t} f_{\lambda}({\bs x}, {\bs k}, t) + \frac{{\bs k}}{2M} \cdot \nabla_{\bs x} f_{\lambda}({\bs x}, {\bs k}, t)
= \ml{C}_{\lambda}^+({\bs x}, {\bs k}, t)[f^{(8)}] - \ml{C}_{\lambda}^-({\bs x}, {\bs k}, t)[f_{\lambda}]  \,,
\end{align}
where the collision kernels
\begin{eqnarray}
\ml{C}_{\lambda}^+({\bs x}, {\bs k}, t)[f^{(8)}] &=& \int\frac{\diff^3 k'}{(2\pi)^3}e^{i {\bs k}'\cdot {\bs x} } \Big\langle {\bs k}+\frac{{\bs k}'}{2},\lambda\Big| \partial_t \widetilde{\mathcal{C}}^{+}(t)\Big|{\bs k}-\frac{{\bs k}'}{2},\lambda\Big\rangle \,, \nn\\
\ml{C}_{\lambda}^-({\bs x}, {\bs k}, t)[f_{\lambda}] &=& \int\frac{\diff^3 k'}{(2\pi)^3}e^{i {\bs k}'\cdot {\bs x} } \Big\langle {\bs k}+\frac{{\bs k}'}{2},\lambda\Big| \partial_t \widetilde{\mathcal{C}}^{-}(t)\Big|{\bs k}-\frac{{\bs k}'}{2},\lambda\Big\rangle \,,
\end{eqnarray}
explicitly depend on polarization. The superscript $(8)$ and the absence of the subscript $\lambda$ in $f^{(8)}$ indicate that the distribution is for unbound heavy quark-antiquark pairs in the color-octet spin-singlet state. The derivation gives the kinetic equation at $t=0$, the initial time (that we previously labeled as $t_i$). As explained earlier, we have generalized the derivation to an arbitrary initial time $t$.

The spin alignment observable is defined by the diagonal components of the normalized spin density matrix (which is related to but different from the system density matrix $\rho_S$ in the spin space)
\begin{equation}
	\rho_{\lambda\lambda}(\bm k)=\frac{\int {\rm d}\Sigma_{x}\cdot k \,f_{\lambda}({\bs x}, {\bs k}, t)}{\int {\rm d}\Sigma_{x}\cdot k\,\sum_{\lambda'=\pm 1,0}\,f_{\lambda'}({\bs x}, {\bs k}, t)} \,, \label{diag_spindensity}
\end{equation}
where ${\rm d}\Sigma_{x\mu}$ represents an element on a freeze-out hypersurface and $k^\mu=(k^0,{\bs k})$ with $k^0=\sqrt{{\bs k}^2+m_\lambda^2}$ and $m_{\lambda}$ being the mass of the vector meson. Nontrivial spin alignment is then characterized by the deviation of $\rho_{00}$ from $\frac{1}{3}$.
When $f_{0}=f_{+}=f_{-}$, $\rho_{00}=\frac{1}{3}$, there is no spin alignment. On the other hand, the quantity $\rho_{++}-\rho_{--}$ is proportional to the spin polarization of the vector meson. The dynamical spin alignment or spin polarization of vector quarkonia can be induced by polarization-dependent collision terms in the Boltzmann transport equation. Our primary goal of this paper is to derive the explicit form of $\ml{C}_{\lambda}^{\pm}$ defined above for the polarization-dependent case, which extends the previous work on the polarization-independent case~\cite{Yao:2020eqy}.

\subsection{Quantum Brownian motion limit}
\label{sect:qbm}
The quantum Brownian motion limit is valid when the environment correlation time is much smaller than the system relaxation time and the intrinsic timescale of the system, i.e., $\tau_R\gg \tau_E$, $\tau_S\gg\tau_E$. This hierarchy of timescales allows one to expand $O^{(S)}_\alpha(t_1)$ in Eq.~\eqref{eqn:pre_lindblad} as
\begin{align}
\label{eqn:Oexpand}
O^{(S)}_\alpha(t_1) = e^{iH_St_1} O^{(S)}_\alpha(0) e^{-iH_St_1} = O^{(S)}_\alpha(0) + i[H_S, O^{(S)}_\alpha(0)]t_1 + \ml{O}\big((H_St_1)^2\big) \,,
\end{align}
and only keep the first few terms. The expansion corresponds to a high-temperature expansion in terms of $\frac{H_S}{T}$ and is justified when $\tau_S\gg \tau_E$ since $\tau_S\sim\frac{1}{H_S}$ and $\tau_E\sim\frac{1}{T}$. If one keeps terms in Eq.~\eqref{eqn:Oexpand} up to leading order (LO) or next-to-leading order (NLO), the resulting evolution equation of $\rho_S(t)$ can be written as a Lindblad equation, which in the Schr\"odinger picture is given by
\begin{align}
\frac{{\rm d}\rho_S(t)}{{\rm d}t} = -i[H_S+\Delta H_S, \rho_S(t)] + \sum_{\alpha,\beta} D_{\alpha\beta}(\omega=0) \left( L_{\beta} \rho_S(t) L_\alpha^\dagger -\frac{1}{2}\left\{ L_\alpha^\dagger L_{\beta}, \rho_S(t) \right\}\right) \,,
\end{align}
where we have used the time translation invariance of the environment correlator and defined
\begin{eqnarray}
D_{\alpha\beta}(t_1,t_2)=D_{\alpha\beta}(t_1-t_2)=\int_{-\infty}^{\infty} \frac{{\rm d}\omega}{2\pi}\, e^{-i\omega (t_1-t_2)} D_{\alpha\beta}(\omega) \,.
\end{eqnarray}
At LO in $\frac{H_S}{T}$, the correction to the system Hamiltonian and the Lindblad operators is given by
\begin{align}
\Delta H_S \xrightarrow{\rm LO} \frac{1}{2}\sum_{\alpha,\beta} \Sigma_{\alpha\beta}(\omega=0) L_\alpha^\dagger L_\beta \,,\qquad
L_\alpha \xrightarrow{\rm LO} O^{(S)}_\alpha(0) \,,\qquad L_\alpha^\dagger \xrightarrow{\rm LO} O^{(S)}_\alpha(0) \,,
\end{align} 
where the argument of $O_\alpha^{(S)}$ describes time $t$, which will be suppressed in the following. The new environment correlation function $\Sigma_{\alpha\beta}$ is defined by
\begin{align}
\Sigma_{\alpha\beta}(\omega) = -i \int_{-\infty}^{\infty} {\rm d}t\, e^{i\omega t} {\rm sign}(t) D_{\alpha\beta}(t) \,.
\end{align}
At NLO, they are
\begin{align}
\label{eqn:Lindblad_NLO}
\Delta H_S & \xrightarrow{\rm NLO} \frac{1}{4}\sum_{\alpha,\beta} \frac{\partial \Sigma_{\alpha\beta}(\omega=0)}{\partial\omega} \left( \left[H_S, O^{(S)}_\alpha \right]O^{(S)}_\beta - O^{(S)}_\alpha \left[H_S, O^{(S)}_\beta \right]\right) \,, \nn\\
L_\alpha & \xrightarrow{{\rm LO}+{\rm NLO}} O^{(S)}_\alpha -\frac{1}{2}\sum_{\beta,\gamma} D^{-1}_{\alpha\beta}(\omega=0) \frac{\partial D_{\beta\gamma}(\omega=0)}{\partial\omega} \left[ H_S, O_\gamma^{(S)} \right] \,, \nn\\
L_\alpha^\dagger & \xrightarrow{{\rm LO}+{\rm NLO}} O^{(S)}_\alpha + \frac{1}{2}\sum_{\beta,\gamma} \left[ H_S, O_\gamma^{(S)} \right] \frac{\partial D_{\gamma\beta}(\omega=0)}{\partial\omega} D^{-1}_{\beta\alpha}(\omega=0) \,,
\end{align}
respectively. When the system is polarization dependent, one can study the time evolution of $\langle \lambda | \rho_S(t) | \lambda \rangle$ and calculate the spin alignment as in Eq.~\eqref{diag_spindensity}.

\section{Spin-Dependent Interaction in pNRQCD}
\label{sect:3}
In this section, we explain how to write down the spin-dependent interaction in pNRQCD. This has been done in Refs.~\cite{Brambilla:2003nt,Brambilla:2004jw}, and the leading thermal correction has been studied in Ref.~\cite{Brambilla:2011mk}. Here we review the derivation of the Lagrangian in Secs.~\ref{sect:3.1} and~\ref{sect:3.2}, since most previous studies that used pNRQCD in the open quantum system approach did not consider the spin-dependent chromomagnetic interaction. We will also discuss the projection onto different spin states, which is new and makes the spin dependence of states manifest.

Our starting point is nonrelativistic QCD (NRQCD) with magnetic interactions~\cite{Bodwin:1994jh}
\begin{align}
\label{eqn:Lnrqcd}
\mathcal{L}_{\rm NRQCD} &= \mathcal{L}_{0} + \mathcal{L}_{1} \,, \nn\\
\mathcal{L}_0 &= \psi^\dagger \left(iD_0 + \frac{{\boldsymbol D}^2}{2M}  \right) \psi + \chi^\dagger \left(iD_0 - \frac{{\boldsymbol D}^2}{2M} \right) \chi \,, \nn\\
\mathcal{L}_1 &= \frac{c_1}{8M^3}\left( \psi^\dagger ({\boldsymbol D}^2)^2 \psi - \chi^\dagger ({\boldsymbol D}^2)^2 \chi \right) \nn\\
& + \frac{c_2}{8M^2}\left(  \psi^\dagger ({\boldsymbol D}\cdot g{\boldsymbol E} - g{\boldsymbol E} \cdot {\boldsymbol D} ) \psi + \chi^\dagger ({\boldsymbol D}\cdot g{\boldsymbol E} - g{\boldsymbol E} \cdot {\boldsymbol D} ) \chi \right) \nn\\
& + \frac{c_3}{8M^2}\left( \psi^\dagger \left( i{\boldsymbol D}\times g{\boldsymbol E} - g{\boldsymbol E} \times i{\boldsymbol D} \right) \cdot {\boldsymbol \sigma} \psi + \chi^\dagger \left( i{\boldsymbol D}\times g{\boldsymbol E} - g{\boldsymbol E} \times i{\boldsymbol D} \right) \cdot {\boldsymbol \sigma} \chi \right) \nn\\
& + \frac{c_4}{2M} \left( \psi^\dagger g{\boldsymbol B}\cdot {\boldsymbol \sigma} \psi - \chi^\dagger g{\boldsymbol B}\cdot {\boldsymbol \sigma} \chi \right) \,,
\end{align}
where $\psi$ ($\chi^\dagger$) denotes the heavy quark (antiquark) annihilation operator and the heavy quark mass is denoted as $M$. The covariant derivatives are defined as $D_\mu = \partial_\mu - igA_\mu$ and act on everything on its right. The chromoelectric and chromomagnetic fields are ${\bs E}$ and ${\bs B}$, and $\bs \sigma$ denotes $(\sigma_x, \sigma_y, \sigma_z)$. The heavy quark fields $\psi$ and $\chi$ carry both color and spin indices, more explicitly, $\psi_i^s$, $\chi_i^s$, where $i$ denotes the color index in the fundamental representation, while $s$ denotes the spin index\footnote{Since $\chi$ creates a heavy antiquark rather than annihilates, it carries the fundamental index rather than the anti-fundamental one.}. The coefficients $c_i$ in $\ml{L}_1$ are Wilson coefficients\footnote{We follow the convention of Ref.~\cite{Bodwin:1994jh} for the subscripts.}. They can be calculated by matching QCD and NRQCD at the scale $M$ order by order in $\alpha_s(\mu=M)$. At leading order, they are $c_i=1$.

In one of the early works on NRQCD~\cite{Bodwin:1994jh}, the power counting rule is obtained by assuming the heavy quark-antiquark pair in quarkonium interacts via a Coulomb potential. Terms in the Lagrangian are organized according to their relative velocity $v$ scaling properties. It is assumed $v\ll1$. Terms in $\ml{L}_0$ scale as $v^2$, while those in $\ml{L}_1$ scale as $v^4$. However, as several following studies noted~\cite{Beneke:1997av,Fleming:2000ib}, a universal $v$ scaling rule cannot be assigned to all the operators in NRQCD. Below the cutoff of NRQCD, three scales exist in vacuum: $Mv, Mv^2, \Lambda_{\rm QCD}$. When $\Lambda_{\rm QCD}$ is non-negligible compared with $Mv$, the power counting becomes more complicated, which is probably the case for charmonium, in particular, the excited states in which the heavy quarks are more nonrelativistic but less Coulombic compared with the ground state. Reference~\cite{Brambilla:2004jw} proposed a new power counting rule where the NRQCD operators are organized by powers of $\frac{1}{M}$. In this new power counting, the leading term is the $iD_0$ term in $\ml{L}_0$ and the leading correction contains the $\frac{{\bs D}^2}{2M}$ term in $\ml{L}_0$ and the magnetic term in $\ml{L}_1$ (the last term with the Wilson coefficient $c_4$). 

Another way of seeing that we only need the magnetic correction term is to note that the $c_1$ and $c_2$ terms in $\ml{L}_1$ are spin independent and thus do not contribute to the physical process of our interest here (polarization change induced by spin flips). Furthermore, we are interested in spin-dependent interactions in a medium, where the typical energy-momentum scale of the gauge field is given by the temperature $T$. So we expect $g{\bs E}\sim gT^2$ and $g{\bs B}\sim gT^2$ instead of $Mv^3$ and $Mv^4$ in the original NRQCD power counting, respectively. Then we see the $c_3$ term is suppressed with respect to the $c_4$ term, by $\frac{T}{M}$ if ${\bs D}\sim T$ or $v$ if ${\bs D}\sim Mv$. Therefore, we only need the $c_4$ term for our leading-order consideration. The NRQCD Lagrangian we are going to use in the following construction is shortened as
\begin{align}
\mathcal{L}_{\rm NRQCD} = \psi^\dagger \left(iD_0 + \frac{{\boldsymbol D}^2}{2M} + c_4 \frac{g{\boldsymbol B}\cdot {\boldsymbol \sigma}}{2M} \right) \psi + \chi^\dagger \left(iD_0 - \frac{{\boldsymbol D}^2}{2M} - c_4 \frac{g{\boldsymbol B}\cdot {\boldsymbol \sigma}}{2M}\right) \chi \,.
\end{align}

The construction of pNRQCD starts with a heavy quark-antiquark composite field
\be
\Phi({\boldsymbol x}_1, {\boldsymbol x}_2, t) = \psi({\boldsymbol x}_1, t) \otimes \chi^\dagger({\boldsymbol x}_2, t)\,,
\ee
which annihilates a pair of heavy quark and antiquark at position ${\bs x}_1$ and ${\bs x}_2$, respectively. We want to emphasize that both $\psi$ and $\chi^\dagger$ carry color and spin indices. In the following, we will work under the hierarchy of energy scales $M\gg Mv \gg Mv^2,T,\Lambda_{\rm QCD}$. We will first briefly review how to obtain the spin-independent pNRQCD, which has been used in many previous studies. Then we will explain how to obtain the spin-dependent one. The final Lagrangian that we will obtain is valid up to the order $\frac{r}{M^0}=r$ and $\frac{r^0}{M}=\frac{1}{M}$, where $r$ denotes the typical relative distance between the $Q\bar{Q}$ pair.

\subsection{Spin-independent case}
\label{sect:3.1}
Here we will omit the spin indices of the heavy quark and antiquark fields and write the color indices out explicitly
\be
\Phi_{ij}({\boldsymbol x}_1, {\boldsymbol x}_2, t) = \psi_i({\boldsymbol x}_1, t) \chi^\dagger_j({\boldsymbol x}_2, t)\,,
\ee
where we have omitted the $\otimes$ symbol since the meaning of the expression is clear after the color indices are spelled out explicitly. From the $\ml{L}_0$ term in Eq.~\eqref{eqn:Lnrqcd}, we can easily write down the Lagrangian for the composite field $\Phi({\boldsymbol x}_1, {\boldsymbol x}_2, t)$, which is
\begin{align}
\label{eqn:LpNRQCD0}
\mathcal{L}({\boldsymbol x}_1,{\boldsymbol x}_2,t) =  {\rm Tr}_c \left[ \Phi^\dagger({\boldsymbol x}_1, {\boldsymbol x}_2, t) \left( iD_0 + \frac{{\bs D}_{{\bs x}_1}^2}{2M} + \frac{{\bs D}_{{\bs x}_2}^{2}}{2M} \right) \Phi({\boldsymbol x}_1, {\boldsymbol x}_2, t)\right] \,,
\end{align}
where the trace ${\rm Tr}_c$ is over color indices and the covariant derivative is 
\begin{align}
\label{eqn:L_phi}
iD_0 \Phi({\boldsymbol x}_1, {\boldsymbol x}_2, t) = 
i\partial_0 \Phi({\boldsymbol x}_1, {\boldsymbol x}_2, t) + gA_0({\boldsymbol x}_1,t)\Phi({\boldsymbol x}_1, {\boldsymbol x}_2, t) - \Phi({\boldsymbol x}_1, {\boldsymbol x}_2, t) gA_0({\boldsymbol x}_2,t) \,.
\end{align}
Then one decomposes the composite field as a color-singlet field $S$ and a color-octet field $O^a$
\begin{align}
\label{eqn:phi_ij}
\Phi_{ij}({\boldsymbol x}_1, {\boldsymbol x}_2, t) = U_{ik}({\bs x}_1, {\bs R}) \left( \frac{\delta_{k\ell}}{\sqrt{N_c}} S({\bs R},{\bs r},t) + \frac{T^a_{k\ell}}{\sqrt{T_F}} O^a({\bs R},{\bs r},t) \right) U_{\ell j}({\bs R}, {\bs x}_2)\,,
\end{align}
where ${\bs R}=({\bs x}_1+{\bs x}_2)/2$ and ${\bs r}={\bs x}_1-{\bs x}_2$ are the center-of-mass (cm) and relative positions, respectively, and 
\begin{align}
U({\bs x}, {\bs y}) \equiv \mathcal{P}\exp\left( ig\int_{\bs y}^{\bs x} {\bs A}({\bs z}) \cdot \diff{\bs z} \right) \,,
\end{align}
denotes a spatial Wilson line with a straight path from $\bs y$ to $\bs x$. $\ml{P}$ denotes path ordering. Gauge fields are matrix variables, i.e., $A_i = A_i^aT^a$, where $T^a$ is the generator of the SU($N_c$) group and is normalized as ${\rm Tr}_c(T^aT^b) = T_F\delta^{ab}$ with $T_F=1/2$.

Plugging Eq.~\eqref{eqn:phi_ij} into Eq.~\eqref{eqn:LpNRQCD0} and performing a multipole expansion in terms of ${\bs r}$ and a matching calculation at the scale $Mv$, one finds
\begin{align}
\label{eqn:L_free+E}
\ml{L} &= S^\dagger({\boldsymbol R}, {\boldsymbol r},t) (i\partial_0 -\ml{H}_s) S({\boldsymbol R}, {\boldsymbol r},t) + O^{a\dagger}({\boldsymbol R}, {\boldsymbol r},t) (iD_0 -\ml{H}_o) O^a({\boldsymbol R}, {\boldsymbol r},t) \nn\\ 
&+ V_A \sqrt{\frac{T_F}{N_c}} \left( S^\dagger({\boldsymbol R}, {\boldsymbol r},t) {\bs r}\cdot g{\bs E}^a({\bs R},t) O^a({\boldsymbol R}, {\boldsymbol r},t) + O^{a\dagger}({\boldsymbol R}, {\boldsymbol r},t) {\bs r}\cdot g{\bs E}^a({\bs R},t) S({\boldsymbol R}, {\boldsymbol r},t) \right) \nn\\
&+ \frac{1}{2} V_B d^{abc} O^{a\dagger}({\boldsymbol R}, {\boldsymbol r},t) {\bs r}\cdot g{\bs E}^b({\bs R},t) O^c({\boldsymbol R}, {\boldsymbol r},t) + \mathcal{O}(r^2) \,,
\end{align}
where $D_\mu O^a \equiv \partial_\mu O^a +g f^{abc} A_\mu^b O^c$ and the color structure constants are defined as $T_Ff^{abc} \equiv -i {\rm Tr}_c(T^a [T^b,T^c])$ and $d^{abc}\equiv 2{\rm Tr}_c(T^a \{T^b,T^c\}) $. The Wilson coefficients $V_A$ and $V_B$ can be calculated at the scale $\mu=Mv$ from the matching and at leading order they are equal to unity $V_A=V_B=1$. The singlet and octet Hamiltonians are 
\begin{align}
\ml{H}_s = -\frac{{\bs \nabla}_{\rm cm}^2}{4M} - \frac{{\bs \nabla}_{\rm rel}^2}{M} + V_s(r) \,,\qquad \ml{H}_o = -\frac{{\bs D}_{\rm cm}^2}{4M} - \frac{{\bs \nabla}_{\rm rel}^2}{M} + V_o(r) \,,
\end{align}
in which the potentials are obtained by adding a potential term $\ml{L}_{\rm pot} = {\rm Tr}_c(\Phi_{ij}^\dagger V \Phi_{ij})$ to the pNRQCD Lagrangian~\eqref{eqn:LpNRQCD0} where $V$ is nonlocal in $r$ and matching with NRQCD order by order in $\frac{1}{M}$ and $\alpha_s(\mu=Mv)$~\cite{Brambilla:2004jw}. The potential term accounts for the physical effects from potential gluons whose energy and momentum scale as $Mv^2$ and $Mv$ respectively. At leading order, we have Coulomb potentials
\begin{align}
\label{eqn:potentials}
V_s(r) = -C_F\frac{\alpha_s}{r} \,,\qquad V_o(r) = \frac{1}{2N_c}\frac{\alpha_s}{r} \,,
\end{align}
where $C_F\equiv (N_c^2-1)/(2N_c)$.

\subsection{Spin-dependent case}
\label{sect:3.2}
We now include spin indices in the composite field construction
\be
\Phi_{ij}^{s_1s_2}({\boldsymbol x}_1, {\boldsymbol x}_2, t) = \psi_i^{s_1}({\boldsymbol x}_1, t) \chi^{\dagger s_2}_j({\boldsymbol x}_2, t)\,. 
\ee
From the magnetic Lagrangian for $\psi$ and $\chi^\dagger$, we can write down the magnetic part of the Lagrangian for $\Phi_{ij}^{s_1s_2}$
\begin{align}
\mathcal{L}_b({\boldsymbol x}_1,{\boldsymbol x}_2,t) &= \frac{c_4}{2M} {\rm Tr}_s {\rm Tr}_c \left[ \Phi^\dagger({\boldsymbol x}_1, {\boldsymbol x}_2, t) g{\boldsymbol B}({\boldsymbol x}_1,t)\cdot {\boldsymbol \sigma}_1 \Phi({\boldsymbol x}_1, {\boldsymbol x}_2, t) \right. \nn\\
&\qquad\qquad\quad\ \ \left. + \Phi^\dagger({\boldsymbol x}_1, {\boldsymbol x}_2, t) \Phi({\boldsymbol x}_1, {\boldsymbol x}_2, t) g{\boldsymbol B}({\boldsymbol x}_2,t)\cdot {\boldsymbol \sigma}_2 \right] \,,
\end{align}
where $B_i=B_i^aT^a$ and ${\rm Tr}_s$ denotes trace over the spin indices. ${\bs \sigma}_1$ and ${\bs \sigma}_2$ act on the heavy quark and antiquark field respectively. More precisely, we can write
\be
{\bs \sigma}_1 = {\bs \sigma} \otimes { I} \,, \qquad {\bs \sigma}_2 = {I} \otimes {\bs \sigma} \,,
\ee
where ${I}$ is an identity operator on spin. 
The $iD_0+\frac{{\bs D}^2}{2M}$ part of the Lagrangian for $\Phi$ in the spin-dependent case is the same as Eq.~\eqref{eqn:LpNRQCD0} with ${\Tr}_s$ added. Decomposing the composite field in terms of color-singlet and color-octet fields and performing a multipole expansion as done in the spin-independent case, we find
\begin{align}
\label{eqn:LpNRQCD_s}
& \mathcal{L}_b = \frac{c_4}{2M} \times \nn\\
& \bigg( V_{A}^s \sqrt{\frac{T_F}{N_c}} {\rm Tr}_s\left[ S^\dagger({\bs R},{\bs r},t) g{\boldsymbol B}^a({\boldsymbol R},t)\cdot {\boldsymbol \sigma}_1 O^a({\bs R},{\bs r},t) + O^{a \dagger}({\bs R},{\bs r},t) g{\boldsymbol B}^a({\boldsymbol R},t)\cdot {\boldsymbol \sigma}_1 S({\bs R},{\bs r},t) \right. \nn\\
& \qquad\qquad\quad \left. +\, S^\dagger({\bs R},{\bs r},t) O^a({\bs R},{\bs r},t) g{\boldsymbol B}^a({\boldsymbol R},t)\cdot {\boldsymbol \sigma}_2 + O^{a \dagger}({\bs R},{\bs r},t)  S({\bs R},{\bs r},t) g{\boldsymbol B}^a({\boldsymbol R},t)\cdot {\boldsymbol \sigma}_2
\right] + \nn\\
& V_{B}^s\frac{{\rm Tr}_c(T^aT^bT^c)}{T_F} {\rm Tr}_s \left[  O^{a\dagger}({\bs R},{\bs r},t) \left( g{\boldsymbol B}^b({\boldsymbol R},t)\cdot {\boldsymbol \sigma}_1 O^c({\bs R},{\bs r},t) + O^b({\bs R},{\bs r},t) g{\boldsymbol B}^c({\boldsymbol R},t)\cdot {\boldsymbol \sigma}_2 \right) \right] \bigg)\nn\\
& + \mathcal{O}(r) \,,
\end{align}
where $V_A^s$ and $V_B^s$ are new Wilson coefficients from matching pNRQCD with NRQCD at the scale $\mu=Mv$. They can be calculated order by order in $\alpha_s(\mu=Mv)$. At leading order they are $V_A^s=V_B^s=1$. Here we mainly use the equation of motion to find the effective field theory, which is equivalent to a tree-level matching condition. In general, operators of the form 
\begin{align}
& \frac{1}{M} {\rm Tr}_c(T^aT^bT^c){\rm Tr}_s\Big[ O^{a\dagger} \sigma_{1i} O^b gB^c_i + O^{a\dagger} gB^b_i O^c \sigma_{2i} \Big] \,,\nn\\
& \frac{1}{M} {\rm Tr}_c(T^aT^bT^c) \hat{r}_i \hat{r}_j{\rm Tr}_s\Big[ O^{a\dagger} \sigma_{1i} O^b gB^c_j + O^{a\dagger} gB^b_i O^c \sigma_{2j} \Big] \,, \nn\\
& \frac{1}{M} \hat{r}_i \hat{r}_j{\rm Tr}_s\Big[ O^{a\dagger} gB^a_i \sigma_{1j}  S  + O^{a\dagger} S gB^a_i \sigma_{2j} + h.c. \Big] \,,
\end{align}
are gauge invariant, respect the nonrelativistic rotation symmetry, and have the same power counting as the terms shown in Eq.~\eqref{eqn:LpNRQCD_s}. Thus they
can appear in the Lagrangian of the effective field theory at the current expansion order we are considering. The Wilson coefficients of these operators vanish at tree level so we will not consider them in this work~\cite{Brambilla:2003nt}. In the matching procedure, the potentials in Eq.~\eqref{eqn:potentials} also acquire spin-dependent corrections such as hyperfine splitting, which can be found in Ref.~\cite{Brambilla:2004jw} and should be included in practical calculations if one wants to account for the mass splitting between the vector meson $J/\psi$ and the pseudoscalar $\eta_c$. We note that the hyperfine splitting occurs at order $\frac{1}{M^2}$, while we focus on the chromomagnetic effects at order $\frac{1}{M}$ here.

The final step of the construction is to project onto particular spin states. The heavy antiquark annihilation field $\chi^\dagger$ transforms in the antifundamental representation, but since SU(2) for the spin rotation is pseudoreal, the antifundamental representation can be made to transform in the same way as the fundamental representation by a basis change given by $-i\sigma_y$. Then the textbook construction of spin singlet ($d=1$) and triplet ($d=3$) can be directly applied. More specifically, we consider the projection
\be
P_d^{s_1s_2} \psi^{s_1} (-i\sigma_y\chi^\dagger)^{s_2} = -iP_d^{s_1s} \sigma_y^{ss_2}\psi^{s_1} \chi^{\dagger s_2} \,.
\ee
For the spin-singlet state, we take
\begin{align}
P_1^{s_1s_2} = \frac{1}{\sqrt{2}}(\delta_{s_1\uparrow}\delta_{s_2\downarrow} - \delta_{s_1\downarrow}\delta_{s_2\uparrow} ) = \frac{1}{\sqrt{2}}\begin{pmatrix}
0 & 1\\
-1 & 0 \end{pmatrix} \,,
\end{align}
where we have written it in a matrix format. Similarly, for spin-triplet states, we have
\begin{align}
P_{3+}^{s_1s_2} &= \delta_{s_1\uparrow}\delta_{s_2\uparrow}  = \begin{pmatrix}
1 & 0\\
0 & 0 \end{pmatrix} \,, \qquad P_{3-}^{s_1s_2} = \delta_{s_1\downarrow}\delta_{s_2\downarrow} = \begin{pmatrix}
0 & 0\\
0 & 1 \end{pmatrix} \,,\nn\\
P_{30}^{s_1s_2} &= \frac{1}{\sqrt{2}}(\delta_{s_1\uparrow}\delta_{s_2\downarrow} + \delta_{s_1\downarrow}\delta_{s_2\uparrow} ) = \frac{1}{\sqrt{2}}\begin{pmatrix}
0 & 1\\
1 & 0 \end{pmatrix} \,.
\end{align}
Including the $-i\sigma_y$, we find
\begin{align}
&-iP_1\sigma_y = \frac{1}{\sqrt{2}} \begin{pmatrix}
1 & 0\\
0 & 1 \end{pmatrix} = \frac{1}{\sqrt{2}}I \,, & 
&-iP_{3+}\sigma_y = - \begin{pmatrix}
0 & 1\\
0 & 0 \end{pmatrix} = - \frac{\sigma_x+i\sigma_y}{2} \,,& \nn\\
&-iP_{30}\sigma_y = \frac{1}{\sqrt{2}} \begin{pmatrix}
1 & 0\\
0 & -1 \end{pmatrix} = \frac{1}{\sqrt{2}} \sigma_z \,, &
&-iP_{3-}\sigma_y = \begin{pmatrix}
0 & 0\\
1 & 0 \end{pmatrix}  = \frac{\sigma_x-i\sigma_y}{2} \,.&
\end{align}
This motivates decomposing the spin part in $S$ and $O^a$ as 
\be
\label{eqn:polarization_decom}
S \to \frac{1}{\sqrt{2}} ( I S_{1} + {\boldsymbol \sigma}\cdot {\bs S}_{3} ) \,,\qquad O^a \to \frac{1}{\sqrt{2}} ( I O^a_{1} + {\boldsymbol \sigma}\cdot {\bs O}^a_{3} ) \,,
\ee
where the subscripts $1$ and $3$ indicate the spin singlet and triplet, respectively. As an example, in the case of radial ground states in charmonium, ${\bm S}_3$ corresponds to $J/\psi$ while $S_1$ represents $\eta_c$.  
When decomposing ${\bs S}_3 = (S_{3 x}, S_{3 y}, S_{3 z})$ and ${\bs O}^a_{3} = (O^a_{3 x}, O^a_{3 y}, O^a_{3 z})$ in terms of creation and annihilation operators, we will include a polarization vector ${\bs \varepsilon}_\lambda = (\varepsilon_{\lambda x}, \varepsilon_{\lambda y}, \varepsilon_{\lambda z})$ in them
\be
{\bs \varepsilon}_{+}=-\frac{1}{\sqrt{2}}(1,i,0) \,,\qquad  {\bs \varepsilon}_{-}=\frac{1}{\sqrt{2}}(1,-i,0) \,, \qquad
{\bs \varepsilon}_{0}=(0,0,1) \,,
\ee
and write 
\begin{align}
S_{3i}({\bs R}, {\bs r}, t) = \sum_{\lambda} {S}_{\lambda i}({\bs R}, {\bs r}, t)\,, \qquad O^a_{3i}({\bs R}, {\bs r}, t) = \sum_{\lambda} O^a_{\lambda i}({\bs R}, {\bs r}, t) \,.
\end{align}

Another way of seeing the Pauli matrix structure in the vector meson and the pseudoscalar is to scrutinize the original Foldy-Wouthuysen-Tani transformation. We let $\Psi_{D}$ be the Dirac spinor in the original QCD Lagrangian. The heavy quark and antiquark fields can be obtained as $\psi\approx\frac{1}{2}(1+\gamma^0)e^{iMt}\Psi_{D}$ and $\chi\approx\frac{1}{2}(1-\gamma^0)e^{-iMt}\Psi_{D}$, respectively, in the Dirac representation of $\gamma$ matrices. We introduce the vector meson and pseudoscalar fields as $\Psi_D\overline{\Psi}_D=\gamma^{\mu}V_{\mu}+\gamma^5{P}$, where we omit the color degrees of freedom and spacetime dependence for brevity. One finds 
\be
\psi \chi^{\dagger}e^{-2iMt}=-\frac{1}{4}\left(\gamma^{\mu}-\gamma^{\mu\dagger}+[\gamma^0,\gamma^{\mu}] \right)V_{\mu}-\frac{1}{2}(1+\gamma^0)\gamma^5{P} \,.
\ee
In the Dirac representation of $\gamma$ matrices, the coefficients of $V_\mu$ and $P$ are given by in block matrix formats 
\begin{align}
-\frac{1}{4}\left(\gamma^{\mu}-\gamma^{\mu\dagger}+[\gamma^0,\gamma^{\mu}] \right) = \begin{pmatrix}
0 & -\delta_{\mu i}\sigma_i\\
0 & 0
\end{pmatrix}\,,\qquad 
-\frac{1}{2}(1+\gamma^0)\gamma^5 =\begin{pmatrix}
0 & -I\\
0 & 0
\end{pmatrix} \,,
\end{align}
where the summation over the repeated index $i$ is Euclidean. The above equation leads to Eq.~\eqref{eqn:polarization_decom} similarly, since the overall minus sign is irrelevant.

Finally, we note that the Lagrangian~\eqref{eqn:LpNRQCD_s} can be rewritten as
\begin{align}
\label{eqn:L_B}
& \mathcal{L}_b = \nn\\
& \frac{c_4}{M} \bigg( V_{A}^s \sqrt{\frac{T_F}{N_c}} \sum_\lambda \left[ S^\dagger_1({\bs R},{\bs r},t) gB_i^a({\boldsymbol R},t) O^a_{\lambda i}({\bs R},{\bs r},t) + O^{a \dagger}_{\lambda i}({\bs R},{\bs r},t) gB_i^a({\boldsymbol R},t) S_1({\bs R},{\bs r},t) \right. \nn\\
& \qquad\qquad\quad\quad\ \,\left. +\, S^\dagger_{\lambda i}({\bs R},{\bs r},t) gB_i^a({\boldsymbol R},t) O^a_1({\bs R},{\bs r},t) + O^{a \dagger}_1({\bs R},{\bs r},t) gB_i^a({\boldsymbol R},t) S_{\lambda i}({\bs R},{\bs r},t) \right] \nn\\
& \quad+ \frac{1}{2} V_{B}^s d^{abc} \sum_\lambda \left[  O^{a\dagger}_1({\bs R},{\bs r},t) g{B}^b_i({\boldsymbol R},t) O^c_{\lambda i}({\bs R},{\bs r},t) + O^{a\dagger}_{\lambda i}({\bs R},{\bs r},t) g{B}^b_i({\boldsymbol R},t) O^c_1({\bs R},{\bs r},t)  \right]  \nn\\
& \quad- \frac{1}{2}V_{B}^s f^{abc}\varepsilon_{ijk} \sum_{\lambda,\lambda'}O_{\lambda i}^{a\dagger}({\bs R},{\bs r},t) gB_j^b({\boldsymbol R},t) O_{\lambda'k}^{c}({\bs R},{\bs r},t) + \mathcal{O}(r) \bigg) \,,
\end{align}
where we have performed the traces over the color and spin indices. The color singlet-octet transition occurs together with the spin singlet-triplet transition. For completeness, we list the chromoelectric part of the Lagrangian, which is the same as the electric interaction part in Eq.~\eqref{eqn:L_free+E} with the trivial spin dependence of the fields added
\begin{align}
\label{eqn:L_E}
&\ml{L}_{e} 	=  V_A \sqrt{\frac{T_F}{N_c}} \left( S_{1}^\dagger({\boldsymbol R}, {\boldsymbol r},t) {\bs r}\cdot g{\bs E}^a({\bs R},t) O^{a}_{1}({\boldsymbol R}, {\boldsymbol r},t) + O_{1}^{a\dagger}({\boldsymbol R}, {\boldsymbol r},t) {\bs r}\cdot g{\bs E}^a({\bs R},t) S_{1}({\boldsymbol R}, {\boldsymbol r},t) \right) \nn\\
&+ \frac{1}{2} V_B d^{abc} O_{1}^{a\dagger}({\boldsymbol R}, {\boldsymbol r},t) {\bs r}\cdot g{\bs E}^b({\bs R},t) O^{c}_{1}({\boldsymbol R}, {\boldsymbol r},t) \, \nn\\
&+V_A \sqrt{\frac{T_F}{N_c}} \!\sum_{\lambda}\!\left(\! S_{\lambda i}^\dagger({\boldsymbol R}, {\boldsymbol r},t) {\bs r}\cdot g{\bs E}^a({\bs R},t) O^{a}_{\lambda i}({\boldsymbol R}, {\boldsymbol r},t) + O_{\lambda i}^{a\dagger}({\boldsymbol R}, {\boldsymbol r},t) {\bs r}\cdot g{\bs E}^a({\bs R},t) S_{\lambda i}({\boldsymbol R}, {\boldsymbol r},t) \!\right) \nn\\
&+ \frac{1}{2} V_B d^{abc} \sum_{\lambda}O_{\lambda i}^{a\dagger}({\boldsymbol R}, {\boldsymbol r},t) {\bs r}\cdot g{\bs E}^b({\bs R},t) O^{c}_{\lambda i}({\boldsymbol R}, {\boldsymbol r},t) + \mathcal{O}(r^2) \,.
\end{align}
The electric interaction does not alter the spin of the composite fields. The electric interaction is at order $\frac{r}{M^0}$ while the magnetic interaction is at order $\frac{r^0}{M}$ in the Lagrangian. The total Lagrangian is valid up to order $(r,\frac{1}{M})$ in the power counting.

\subsection{Canonical quantization}
The total Lagrangian is given by 
$\ml{L}+\ml{L}_b$. $\ml{L}$ can be found in Eq.~\eqref{eqn:L_free+E}, which does not alter the spin of the fields and thus it is the same for all $S_1$, $O^a_1$, ${S}_{\lambda i}$, and ${O}^a_{\lambda i}$. The electric interaction part is contained inside $\ml{L}$ and its trivial polarization dependence can be found in Eq.~\eqref{eqn:L_E} explicitly. The magnetic interaction part is given in Eq.~\eqref{eqn:L_B}, which can alter the spin of the fields. In the canonical quantization procedure\footnote{Here we focus on the Hamiltonian of the color-singlet and color-octet fields and will not discuss the subtlety involved in canonically quantizing the gauge degrees of freedom, which are constrained. In the following expressions of the Hamiltonian, a non-axial gauge has been chosen. The final correlators of chromomagnetic fields to be constructed will be gauge invariant.}, we decompose the Hamiltonian in terms of a free part and interacting parts,
\begin{align}
\label{eqn:Hamiltonians}
&H_0 = \int{\rm d}^3R{\rm d}^3r \left( S^\dagger_1({\bs R},{\bs r}) \ml{H}_s^1 S_1({\bs R},{\bs r}) + S^\dagger_{\lambda i}({\bs R},{\bs r}) \ml{H}_s^\lambda S_{\lambda i}({\bs R},{\bs r}) \right.\nn\\
&\left. \ \qquad\qquad\qquad +\, O^{a \dagger}_1({\bs R},{\bs r}) \ml{H}_{o,f}^1 O^a_1({\bs R},{\bs r}) + O^{a\dagger}_{\lambda i}({\bs R},{\bs r}) \ml{H}_{o,f}^\lambda O^a_{\lambda i}({\bs R},{\bs r}) \right) \,,\nn\\
&H_1 = - \int{\rm d}^3R{\rm d}^3r\, \bigg[ gf^{abc}\Big(i O_1^{a\dagger}({\bs R},{\bs r}) A_0^b({\bs R}) O^c_1({\bs R},{\bs r}) + i O_{\lambda i}^{a\dagger}({\bs R},{\bs r}) A_0^b({\bs R}) O^c_{\lambda i}({\bs R},{\bs r}) \nn\\
& \qquad + \frac{1}{4M}O_1^{a\dagger}({\bs R},{\bs r}) \big[\partial_{R_i} A_i^b({\bs R}) + 2A_i^b({\bs R}) \partial_{R_i} \big] O^c_1({\bs R},{\bs r}) \nn\\
&\qquad + \frac{1}{4M} O_{\lambda j}^{a\dagger}({\bs R},{\bs r}) \big[\partial_{R_i} A_i^b({\bs R}) + 2A_i^b({\bs R}) \partial_{R_i} \big] O^c_{\lambda j}({\bs R},{\bs r}) \Big) \nn\\
& \qquad + \frac{g^2f^{abc}f^{cde}}{4M}\Big( O_1^{a\dagger}({\bs R},{\bs r}) A_i^b({\bs R}) A_i^d({\bs R}) O_1^e({\bs R},{\bs r}) + O_{\lambda j}^{a\dagger}({\bs R},{\bs r}) A_i^b({\bs R}) A_i^d({\bs R}) O_{\lambda j}^e({\bs R},{\bs r}) \Big)  \bigg] \,, \nn\\
&H_2 = -\int{\rm d}^3R{\rm d}^3r \left[ \ml{L}_e({\bs R},{\bs r}) +\ml{L}_b({\bs R},{\bs r}) \right] \,,
\end{align}
where the fields are time independent and $\ml{L}_e$ and $\ml{L}_b$ are given by Eqs.~\eqref{eqn:L_E} and~\eqref{eqn:L_B} with the time dependence removed. In the following, we will mainly focus on the magnetic interaction part and omit the electric interaction in most discussions since our main interest of the paper is the spin-dependent interaction. The Boltzmann equation derived from the electric interaction can be found in Ref.~\cite{Yao:2020eqy}. The $\ml{H}$ terms are given by
\begin{align}
\ml{H}_s^{1/\lambda}  = -\frac{{\bs \nabla}_{\rm cm}^2}{4M} - \frac{{\bs \nabla}_{\rm rel}^2}{M} + V_s^{1/\lambda}(r) \,, \qquad \ml{H}_{o,f}^{1/\lambda} & = -\frac{{\bs \nabla}_{\rm cm}^2}{4M} - \frac{{\bs \nabla}_{\rm rel}^2}{M} + V_o^{1/\lambda}(r) \,,
\end{align}
where we have assumed more generally that the potential can be spin dependent, indicated by the superscripts $1$ and $\lambda$. [The degeneracy between the spin singlet and spin triplet is lifted up due to the hyperfine splitting in general and the spin-orbit coupling for non-$S$-wave states, which first appear in the Lagrangian at order $\frac{1}{M^2}$. At the same order, the potential may contain a term of the form ${\bs r}\times {\bs P}_{\rm cm} \cdot({\bs \sigma}_1-{\bs \sigma}_2)$ beyond tree-level matching~\cite{Brambilla:2003nt}, which can break the energy degeneracy between the three $\lambda=+,-,0$ states.] Part of the $\frac{{\bs D}_{\rm cm}^2}{4M}$ term in $\ml{H}_o$ is treated as an interaction in $H_1$. As a result, only the free part, i.e., $\frac{{\bs \nabla}_{\rm cm}^2}{4M}$ appears in $H_0$, indicated by the subscript $f$. 

In the canonical quantization procedure, the fields are decomposed as creation and annihilation operators which obey the standard commutation relations. In the standard interaction picture defined by $H_0$, we have [we will omit the superscript (int) for simplicity]
\begin{align}
\label{eqn:field_quantize}
S_{1}({\bs R}, {\bs r}, t) &= \int \frac{{\rm d}^3p_{\rm cm}}{(2\pi)^3} \sum_{n\ell} e^{-iEt+i{\bs p}_{\rm cm}\cdot {\bs R}} 
\psi_{n\ell}({\bs r}) a_{n\ell}({\bs p}_{\rm cm}) \nn\\
& + \int \frac{{\rm d}^3p_{\rm cm}}{(2\pi)^3}  \int \frac{{\rm d}^3p_{\rm rel}}{(2\pi)^3} e^{-iE_{p}t+i{\bs p}_{\rm cm}\cdot {\bs R}} \psi_{{\bs p}_{\rm rel}}({\bs r}) b_{{\bs p}_{\rm rel}}({\bs p}_{\rm cm}) \,, \nn\\
S_{\lambda i}({\bs R}, {\bs r}, t) &=  \int \frac{{\rm d}^3p_{\rm cm}}{(2\pi)^3} \sum_{n\ell} e^{-iE^\lambda t+i{\bs p}_{\rm cm}\cdot {\bs R}} \varepsilon_{\lambda i} \, \psi_{n\ell}^\lambda({\bs r}) a_{n\ell}({\bs p}_{\rm cm},\lambda)  \nn\\
& + \int \frac{{\rm d}^3p_{\rm cm}}{(2\pi)^3}  \int \frac{{\rm d}^3p_{\rm rel}}{(2\pi)^3}  e^{-iE^\lambda_{p}t+i{\bs p}_{\rm cm}\cdot {\bs R}} \varepsilon_{\lambda i} \,\psi_{{\bs p}_{\rm rel}}^\lambda({\bs r}) b_{{\bs p}_{\rm rel}}({\bs p}_{\rm cm},\lambda) \,, \nn\\
O^a_1({\bs R}, {\bs r}, t) &= \int \frac{{\rm d}^3p_{\rm cm}}{(2\pi)^3} \int \frac{{\rm d}^3p_{\rm rel}}{(2\pi)^3} e^{-iE_{p}t+i{\bs p}_{\rm cm}\cdot {\bs R}} \Psi_{{\bs p}_{\rm rel}}({\bs r}) c^a_{{\bs p}_{\rm rel}}({\bs p}_{\rm cm}) \,,  \nn\\
O^a_{\lambda i}({\bs R}, {\bs r}, t) &= \int \frac{{\rm d}^3p_{\rm cm}}{(2\pi)^3} \int \frac{{\rm d}^3p_{\rm rel}}{(2\pi)^3} e^{-iE^\lambda_{p}t+i{\bs p}_{\rm cm}\cdot {\bs R}} \varepsilon_{\lambda i} \,\Psi_{{\bs p}_{\rm rel}}^\lambda({\bs r}) c^a_{{\bs p}_{\rm rel}}({\bs p}_{\rm cm},\lambda) \,,
\end{align}
where $E^{(\lambda)}=\frac{{\bs p}^2_{\rm cm}}{4M}+E^{(\lambda)}_{n\ell}$, $E^{(\lambda)}_p = E^{(\lambda)}_{p_{\rm cm}}+E^{(\lambda)}_{p_{\rm rel}} = \frac{{\bs p}^2_{\rm cm}}{4M} + \frac{{\bs p}^2_{\rm rel}}{M}$ and $a,b,c$ denote the standard annihilation operators with the corresponding quantum numbers. They satisfy the nonrelativistic commutation relation, e.g.,
\begin{align}
[a_{n_1\ell_1}({\bs p}_{1{\rm cm}},\lambda_1), a_{n_2\ell_2}^\dagger({\bs p}_{2{\rm cm}},\lambda_2)] &= (2\pi)^3\delta^3({\bs p}_{1{\rm cm}} - {\bs p}_{2{\rm cm}}) \delta_{\lambda_1\lambda_2} \delta_{n_1n_2}\delta_{\ell_1\ell_2} \,, \nn\\
[b_{{\bs p}_{1{\rm rel}}}({\bs p}_{1{\rm cm}},\lambda_1), b^\dagger_{{\bs p}_{2{\rm rel}}}({\bs p}_{2{\rm cm}},\lambda_2)] &= (2\pi)^6\delta^3({\bs p}_{1{\rm cm}} - {\bs p}_{2{\rm cm}}) \delta^3({\bs p}_{1{\rm rel}} - {\bs p}_{2{\rm rel}}) \delta_{\lambda_1\lambda_2} \,, \nn\\
[c^{a_1}_{{\bs p}_{1{\rm rel}}}({\bs p}_{1{\rm cm}}), c^{a_2\dagger}_{{\bs p}_{2{\rm rel}}}({\bs p}_{2{\rm cm}})] &= (2\pi)^6\delta^3({\bs p}_{1{\rm cm}} - {\bs p}_{2{\rm cm}}) \delta^3({\bs p}_{1{\rm rel}} - {\bs p}_{2{\rm rel}}) \delta^{a_1a_2} \,.
\end{align}
Other commutation relations can be similarly written out. 
For the relative motion, $\psi_{n\ell}$ denotes a bound state wave function in the color-singlet, spin-singlet channel with the quantum number $n\ell$ ($n$ stands for the radial excitation, $\ell$ for the orbital angular momentum and the third component of the orbital angular momentum $m$ is neglected here, which should be averaged in practical calculations due to the energy degeneracy associated with $m$) and a binding energy $E_{n\ell}<0$, while $\psi_{{\bs p}_{\rm rel}}$ represents a scattering wave in the color-singlet, spin-singlet channel whose relative energy is given by $E_{{p}_{\rm rel}}$. Finally, $\Psi_{{\bs p}_{\rm rel}}$ describes a scattering wave in the color-octet, spin-singlet channel. Those wave functions with the superscript $\lambda$ are for the spin-triplet channel. If we neglect the center-of-mass kinetic energies, the corresponding eigenenergies of these wave functions are defined as
\begin{align}
\label{eqn:energies}
&\left( - \frac{{\bs \nabla}_{\rm rel}^2}{M} + V_s^{1}(r) \right) \psi_{n\ell} = E_{n\ell} \psi_{n\ell} \,, &&\left( - \frac{{\bs \nabla}_{\rm rel}^2}{M} + V_s^{\lambda}(r) \right) \psi^\lambda_{n\ell} = E^\lambda_{n\ell} \psi_{n\ell}^\lambda \,, \nn \\
&\left( - \frac{{\bs \nabla}_{\rm rel}^2}{M} + V_s^1(r) \right) \psi_{{\bs p}_{\rm rel}} = E_{{p}_{\rm rel}} \psi_{{\bs p}_{\rm rel}} \,, &&\left( - \frac{{\bs \nabla}_{\rm rel}^2}{M} + V_s^\lambda(r) \right) \psi^\lambda_{{\bs p}_{\rm rel}} = E^\lambda_{{p}_{\rm rel}} \psi^\lambda_{{\bs p}_{\rm rel}} \,, \nn \\
&\left( - \frac{{\bs \nabla}_{\rm rel}^2}{M} + V_o^1(r) \right) \Psi_{{\bs p}_{\rm rel}} = E_{{p}_{\rm rel}} \Psi_{{\bs p}_{\rm rel}} \,, &&\left( - \frac{{\bs \nabla}_{\rm rel}^2}{M} + V_o^\lambda(r) \right) \Psi^\lambda_{{\bs p}_{\rm rel}} = E^\lambda_{{p}_{\rm rel}} \Psi^\lambda_{{\bs p}_{\rm rel}} \,.
\end{align}
This completes the canonical quantization setup.

\subsection{A special picture for summing $H_1$}
\label{sect:special_picture}
We will treat the magnetic interaction in $H_2$ as a perturbation and expand the unitary evolution operator accordingly since it is suppressed by $\frac{1}{M}$, but we will sum the $iD_0$ and $-\frac{{\bs D}_{\rm cm}^2}{4M}$ interactions in $H_1$ between the octet and the environment to all orders in $g$. This can be achieved by using a special picture defined by $H_1^{({\rm int})}(t) = e^{iH_0t} H_1 e^{-iH_0t}$. More specifically, an arbitrary operator $O$ in the new special picture is defined by
\begin{align}
\widetilde{O}(t) &= \Big( \overline{\ml{T}} e^{i\int_{t_0}^t{\rm d}t H_1^{({\rm int})}(t)} \Big) O^{({\rm int})}(t) \Big(\ml{T}e^{-i\int_{t_0}^t{\rm d}t H_1^{({\rm int})}(t)}\Big) \nn\\
&= \Big( \overline{\ml{T}} e^{i\int_{t_0}^t{\rm d}t H_1^{({\rm int})}(t)} \Big) e^{iH_0t}Oe^{-iH_0t} \Big(\ml{T} e^{-i\int_{t_0}^t{\rm d}t H_1^{({\rm int})}(t)} \Big)\,,
\end{align}
where $\ml{T}$ and $\overline{\ml{T}}$ denote the time-ordering and antiordering operators, respectively, and $t_0$ is the time when $\widetilde{O}(t_0) = O^{({\rm int})}(t_0)$. Since $H_1$ is independent of the singlet field, it will not modify the singlet field in $H_2$. In practice, the gauge field part of $H_1$ should be treated as part of $H_E$ nonperturbatively in this special picture.

The effect of the new special picture induced by the $A_0$ term in $H_1$ is to generate a color rotation given by an adjoint timelike Wilson (straight) line. For example, if we consider an octet state $|{\bs R}, a\rangle$ at center-of-mass position ${\bs R}$, the schematic effect of the new special picture is
\begin{align}
\Big(\ml{T} e^{-i\int_{t_0}^t{\rm d}t H_1^{({\rm int})}(t)} \Big) |{\bs R}, a\rangle \to |{\bs R},b\rangle W^{ba}[({\bs R},t),({\bs R},t_0)]  \,,
\end{align}
where $t_0$ is the time at which the octet state $|{\bs R}, a\rangle$ is created.
$W[({\bs R},t_2), ({\bs R},t_1)]$ denotes an adjoint timelike Wilson (straight) line from $({\bs R},t_1)$ to $({\bs R},t_2)$,
\begin{align}
W[({\bs R},t_2), ({\bs R},t_1)] = \ml{P} \exp\left\{ ig\int_{t_1}^{t_2} {\rm d}t' A_0^{\rm adj}({\bs R},t') \right\} \,,
\end{align}
in which $\ml{P}$ denotes path ordering and $A_0^{\rm adj} = A_0^a T_A^a$ with $(T_A^a)^{bc}=-if^{abc}$. We will be more specific on the effect of $\widetilde{O}^a({\bs R},{\bs r},t)$ in the next section.

The one-gluon and two-gluon interactions in the $-\frac{{\bs D}_{\rm cm}^2}{4M}$ term can be summed over in a diagrammatic approach, as done in Ref.~\cite{Yao:2020eqy}, which gives a spatial adjoint Wilson line at infinite time. This Wilson line appears if one studies physical processes that are differential in the center-of-mass momentum. In our case, we will study inclusive transition processes during which the spin may change. The center-of-mass momentum will be integrated over and thus the spatial Wilson line at infinite time does not appear explicitly.

All in all, the magnetic interaction Hamiltonian in the new special picture can be written as
\begin{align}
& \widetilde{H}_{2b}(t) = -\int{\rm d}^3R {\rm d}^3r \Big(\overline{\ml{T}}e^{i\int_{t_0}^t{\rm d}t H_1^{({\rm int})}(t)} \Big) \nn\\
& \frac{c_4}{M} \bigg( V_{A}^s \sqrt{\frac{T_F}{N_c}} \sum_{\lambda} \left[ S^\dagger_1({\bs R},{\bs r},t) g{B}_i^a({\boldsymbol R},t) {O}^a_{\lambda i}({\bs R},{\bs r},t) + {O}^{a \dagger}_{\lambda i}({\bs R},{\bs r},t) g{B}_i^{a}({\boldsymbol R},t) S_1({\bs R},{\bs r},t) \right. \nn\\
& \qquad\qquad\qquad\quad\!\! \left. +\, S^\dagger_{\lambda i}({\bs R},{\bs r},t) g{B}_i^{a}({\boldsymbol R},t) {O}^a_1({\bs R},{\bs r},t) + {O}^{a \dagger}_1({\bs R},{\bs r},t) g{B}_i^{a}({\boldsymbol R},t) S_{\lambda i}({\bs R},{\bs r},t) \right] \nn\\
& \quad + \frac{1}{2} V_{B}^s d^{abc} \sum_{\lambda} \left[  {O}^{a\dagger}_1({\bs R},{\bs r},t) gB^b_i({\boldsymbol R},t) {O}^{c}_{\lambda i}({\bs R},{\bs r},t) + {O}^{a\dagger}_{\lambda i}({\bs R},{\bs r},t) gB^b_i({\boldsymbol R},t) {O}^{c}_1({\bs R},{\bs r},t)  \right]  \nn\\
& \quad - \frac{1}{2} V_{B}^s f^{abc}\varepsilon_{ijk} \sum_{\lambda,\lambda'} {O}_{\lambda i}^{a\dagger}({\bs R},{\bs r},t) gB^b_j({\boldsymbol R},t) {O}_{\lambda'k}^{c}({\bs R},{\bs r},t) + \mathcal{O}(r) \bigg) \Big(\ml{T}e^{-i\int_{t_0}^t{\rm d}t H_1^{({\rm int})}(t)} \Big) \,.
\end{align}
We are now ready to derive the spin-dependent Boltzmann equation and Lindblad equation for quarkonium.

\section{Boltzmann equation in the quantum optical and semiclassical limits}
\label{sect:Boltzmann_eq}
We shall now construct the collision terms in the Boltzmann equation for the vector quarkonium under the energy scale hierarchy $M\gg Mv \gg Mv^2,T,\Lambda_{\rm QCD}$. The major findings of this section are the dissociation and recombination terms shown in Eqs.~(\ref{eq:C_dissociation}) and~(\ref{eq:C_recombination}), respectively. Since quarkonium is a bound color-singlet state and an octet $Q\bar{Q}$ cannot form a bound state due to the repulsive potential, we will focus on the color singlet-octet transition with the singlet state being bound. Singlet-octet transition with the singlet state being unbound and octet-octet transition can change the spin degrees of freedom of unbound $Q\bar{Q}$ pairs and thus may also influence the quarkonium spin alignment due to recombination effect. In practical applications, one will use coupled Boltzmann equations: the spin degrees of freedom of unbound $Q\bar{Q}$ pairs are described by the Boltzmann equations for open heavy flavors with spin-dependent interactions, while the spins of quarkonium states are described by the Boltzmann equation we are going to derive, with the open heavy flavor distribution as an input.

In order to evaluate the $\widetilde{\ml{C}}^{\pm}$ terms defined in Eq.~\eqref{eqn:Cpm}, we need to specify the eigenstates of the free Hamiltonian $H_0$, which form a complete set of states. They are given by
\begin{align}
&|{\bs k},n\ell \rangle=a^{\dagger}_{n\ell}({\bs k})|0\rangle \,,
&& |{\bs k}, n\ell, \lambda \rangle=a^{\dagger}_{n\ell}({\bs k},\lambda)|0\rangle \,,& \nn\\
&| {\bs p}_{\ma{cm}}, {\bs p}_{\ma{rel}}, a \rangle = c^{a\dagger}_{\bm p_{\rm rel}}({\bm p}_{\rm cm})|0\rangle \,, 
&& | {\bs p}_{\ma{cm}}, {\bs p}_{\ma{rel}}, a, \lambda \rangle=c^{a\dagger}_{\bm p_{\rm rel}}({\bm p}_{\rm cm},\lambda)|0\rangle \,.&
\end{align}
We need to compute the transition amplitudes between these states mediated via $\widetilde{H}_{2b}(t)$. Concretely, we obtain
\begin{align}
\label{eqn:ME_B}
&\langle {\bs k},n\ell | \Big(\!\overline{\ml{T}}e^{i\int_{t_0}^t{\rm d}t H_1^{({\rm int})}(t)} \!\Big) S^\dagger_1({\bs R},{\bs r},t) g{B}_i^a({\bs R},t) {O}^a_{\lambda i}({\bs R},{\bs r},t) \Big(\!\ml{T}e^{-i\int_{t_0}^t{\rm d}t H_1^{({\rm int})}(t)} \!\Big) | {\bs p}_{\ma{cm}}, {\bs p}_{\ma{rel}}, b, \lambda' \rangle \nn\\
&= \delta_{\lambda\lambda'} e^{i(E_{n\ell}-E^{\lambda'}_{{p}_{\rm rel}})t -i({\bs k}-{\bs p}_{\rm cm})\cdot{\bs R}} \psi^*_{n\ell}({\bs r}) \Psi^{\lambda'}_{{\bs p}_{\rm rel}}({\bs r}) gB_i^a({\bs R},t) \varepsilon_{\lambda'i} W^{ab}[({\bs R},t),({\bs R},t_0)] \,, \nn\\[8pt]
&\langle {\bs p}_{\ma{cm}}, {\bs p}_{\ma{rel}}, b, \lambda'| \Big(\!\overline{\ml{T}}e^{i\int_{t_0}^t{\rm d}t H_1^{({\rm int})}(t)} \!\Big) {O}^{a \dagger}_{\lambda i}({\bs R},{\bs r},t) g{B}_i^{a}({\boldsymbol R},t) S_1({\bs R},{\bs r},t) \Big(\!\ml{T}e^{-i\int_{t_0}^t{\rm d}t H_1^{({\rm int})}(t)} \!\Big) | {\bs k}, n\ell \rangle \nn\\
&=\delta_{\lambda\lambda'} e^{-i(E_{n\ell}-E^{\lambda'}_{{p}_{\rm rel}})t + i({\bs k}-{\bs p}_{\rm cm})\cdot{\bs R}} \psi_{n\ell}({\bs r}) \Psi^{\lambda'*}_{{\bs p}_{\rm rel}}({\bs r}) W^{ba}[({\bs R},t_0), ({\bs R},t)]  gB_i^a({\bs R},t) 
\varepsilon_{\lambda'i}^* \,,\nn\\[8pt]
&\langle {\bs k}, n\ell, \lambda' | \Big(\!\overline{\ml{T}}e^{i\int_{t_0}^t{\rm d}t H_1^{({\rm int})}(t)} \!\Big)  S^\dagger_{\lambda i}({\bs R},{\bs r},t) g{B}_i^{a}({\boldsymbol R},t) {O}^a_1({\bs R},{\bs r},t) \Big(\!\ml{T}e^{-i\int_{t_0}^t{\rm d}t H_1^{({\rm int})}(t)} \!\Big) | {\bs p}_{\ma{cm}}, {\bs p}_{\ma{rel}}, b \rangle \nn\\
&= \delta_{\lambda\lambda'} e^{i(E^{\lambda'}_{n\ell}-E_{{p}_{\rm rel}})t -i({\bs k}-{\bs p}_{\rm cm})\cdot{\bs R}} \psi^{\lambda'*}_{n\ell}({\bs r}) \Psi_{{\bs p}_{\rm rel}}({\bs r}) gB_i^a({\bs R},t) \varepsilon^*_{\lambda'i} W^{ab}[({\bs R},t),({\bs R},t_0)] \,, \nn\\[8pt]
&\langle {\bs p}_{\ma{cm}}, {\bs p}_{\ma{rel}}, b | \Big(\!\overline{\ml{T}}e^{i\int_{t_0}^t{\rm d}t H_1^{({\rm int})}(t)} \!\Big) {O}^{a \dagger}_1({\bs R},{\bs r},t) g{B}_i^{a}({\boldsymbol R},t) S_{\lambda i}({\bs R},{\bs r},t) \Big(\!\ml{T}e^{-i\int_{t_0}^t{\rm d}t H_1^{({\rm int})}(t)} \!\Big) |{\bs k}, n\ell, \lambda' \rangle \nn\\
&=\delta_{\lambda\lambda'} e^{-i(E^{\lambda'}_{n\ell}-E_{{p}_{\rm rel}})t +i({\bs k}-{\bs p}_{\rm cm})\cdot{\bs R}} \psi^{\lambda'}_{n\ell}({\bs r}) \Psi^*_{{\bs p}_{\rm rel}}({\bs r}) W^{ba}[({\bs R},t_0),({\bs R},t)] gB_i^a({\bs R},t) \varepsilon_{\lambda'i}  \,,
\end{align}
where $t_0$ labels the time when the relevant octet state is created or annihilated. We have omitted the center-of-mass energy in the time-dependent phase, since in practice we will work in the rest frame of the initial quarkonium or unbound $Q\bar{Q}$ pair, and the recoil momentum of the final state is on the order of $T$, much smaller than $M$, which means the center-of-mass energy is suppressed. From the above equations, we can easily separate the system and environment operators in the interaction Hamiltonian.

In the following computations, we will assume the quarkonium states with different quantum number $n\ell$ are nondegenerate. Moreover, in practice we mainly focus on the transitions that involve the radial ground states such as $J/\psi$ and $\eta_c$, so we can fix $n=1$ and $\ell=0$, but we will make the expressions general.

\subsection{Dissociation}
We first consider the dissociation in the collision term, given by the $\widetilde{\ml{C}}^-$ term in Eq.~\eqref{eqn:Cpm}. We sandwich the $\gamma_{ab,cd}(t) L_{cd}^\dagger L_{ab} \rho_S (0)$ operator between two color-singlet spin-triplet states, which gives $\langle {\bs k}_1, n_1\ell_1, \lambda_1 | \gamma_{ab,cd}(t) L_{cd}^\dagger L_{ab} \rho_S (0)| {\bs k}_2, n_2\ell_2, \lambda_2\rangle$. It yields
$|E_c\rangle=|E_a\rangle$ and $|d \rangle=|{\bs k}_1, n_1\ell_1, \lambda_1\rangle$ ($L_{ab} = |E_a\rangle\langle E_b|$ and $|E_a\rangle$ denotes an eigenstate of the system Hamiltonian). Then, from the structure of the magnetic interaction term in $\gamma_{ab,cd}(t)$ as listed in Eq.~\eqref{eqn:ME_B}, we find $|E_c\rangle=| {\bs p}_{\ma{cm}}, {\bs p}_{\ma{rel}}, a \rangle$ and $|E_b\rangle=|{\bm k}_3, n_3\ell_3, \lambda_3\rangle$, respectively. Following the construction in Refs.~\cite{Yao:2018nmy,Yao:2020eqy} under the Markovian condition, we derive
\begin{align}
\nn
&\sum_{a,b,c,d} \big\langle {\bs k}_1, n_1\ell_1, \lambda_1 \big| \gamma_{ab,cd}(t) L_{cd}^\dagger L_{ab} \rho_S (0)\big| {\bs k}_2, n_2\ell_2, \lambda_2 \big\rangle 
\\
\nn
&=  \frac{(c_4V_A^s)^2T_F}{M^2N_c} \int\frac{\diff^3p_{\ma{cm}}}{(2\pi)^3} \frac{\diff^3p_{\ma{rel}}}{(2\pi)^3} \frac{\diff^3k_{3}}{(2\pi)^3} \int \diff^3R_1 \int \diff^3R_2 \int_{0}^{t} \diff t_1 \int_{0}^{t} \diff t_2 \\ \nn
&\times   e^{i(E_{n_1\ell_1}^{\lambda_1}t_1 - {\bs k}_1\cdot {\bs R}_1 )  -i(E_{p_{\rm rel}}t_1 - {\bs p}_\ma{cm} \cdot {\bs R}_1) } 
e^{-i(E_{n_3\ell_3}^{\lambda_3}t_2 - {\bs k}_3\cdot {\bs R}_2 )  + i(E_{p_{\rm rel}}t_2 - {\bs p}_\ma{cm} \cdot {\bs R}_2) } g_{ij}^{B++}(t_1,t_2,{\bs R}_1,{\bs R}_2) 
\\ 
\label{eqn:step_d1}
&\times \varepsilon^*_{\lambda_1 i}\varepsilon_{\lambda_3j} 
\langle \psi^{\lambda_1}_{n_1\ell_1} |  \Psi_{{\bs p}_\ma{rel}} \rangle  \langle \Psi_{{\bs p}_\ma{rel}} | \psi^{\lambda_3}_{n_3\ell_3} \rangle
\big\langle {\bs k}_3, n_3\ell_3, \lambda_3 \big| \rho_S(0) \big| {\bs k}_2, n_2\ell_2, \lambda_2 \big\rangle\,,
\end{align}
where $E_{p_{\rm rel}}=\frac{{\bs p}^2_{\rm rel}}{M}$, and $E_{n_1\ell_1}^{\lambda_1}$ and $E_{n_3\ell_3}^{\lambda_3}$ are the binding energies of the quarkonium states with the corresponding quantum numbers (which have negative values). Repeated indices are summed over implicitly. The chromomagnetic field correlator is defined by
\begin{align}
g_{ij}^{B++}(t_1,t_2,{\bs R}_1,{\bs R}_2) \equiv &\, \Tr_E\Big\{ gB_i^a({\bs R}_1,t_1) W^{ac}[({\bs R}_1,t_1),({\bs R}_1,\infty)]
W^{cd}[({\bs R}_1,\infty),({\bs R}_2,\infty)] \nn\\
& \qquad \times W^{db}[({\bs R}_2,\infty),({\bs R}_2,t_2)] gB_j^b({\bs R}_2,t_2) \rho_T(0) \Big\}  \,,
\end{align}
where the spatial Wilson line at infinite time is generated by the one- and two-gluon interactions in $\frac{{\bs D}_{\rm cm}^2}{4M}$ as explained in Sec.~\ref{sect:special_picture}. It will disappear in the final collision term for the inclusive dissociation process.
We have set $t_0=\infty$ since the octet states are created after the dissociation and live in the future. However, as we will see later, the $t_0$ dependence in $g_{ij}^{B++}$ will cancel for the inclusive dissociation process. 

For the spin alignment observable, we only need the diagonal part of the spin density matrix. So we assume $\langle \lambda |\rho_S|\lambda'\rangle \propto \delta_{\lambda\lambda'}$ is diagonal and set
$\lambda_1=\lambda_2=\lambda_3=\lambda$. For the distribution function of a particular quarkonium state, we also set $n_1=n_2=n$ and $\ell_1=\ell_2=\ell$. As explained in Sec.~\ref{sect:2}, when the relaxation rate of the system is much larger than the expansion rate of the QGP, we can treat the environment correlator as translationally invariant in time. Furthermore, since we mainly consider heavy quark pairs moving slowly in the QGP, we will assume the environment correlator is also translationally invariant in space. We may now introduce $\delta t=t_1-t_2$, $\delta {\bs R}={\bs R}_1-{\bs R}_2$ and the Fourier transform of $g^{B++}_{ij}(t_1, t_2, {\bs R}_1, {\bs R}_2) = g^{B++}_{ij}(t_1-t_2, {\bs R}_1-{\bs R}_2)$ as
\begin{eqnarray}
\tilde{g}^{B++}_{ij}(q)=\int\diff\delta t \diff^3\delta R \, e^{iq_0\delta t-i\bs q\cdot\delta\bs R}g^{B++}_{ij}(t_1, t_2, {\bs R}_1, {\bs R}_2) \,.
\end{eqnarray}
Under the Markovian approximation, the upper limit of the time integral in Eq.~\eqref{eqn:step_d1} can be taken to be $t=\infty$ and then the two time integrals give a $\delta$ function for energy conservation, multiplied by the time length $t$. The energy conservation gives $n_1=n_3=n$ and $\ell_1=\ell_3=\ell$ due to the nondegeneracy assumed. Performing the remaining multiple integrals leads to
\be\nn\label{eqn:step_d2}
&&\sum_{a,b,c,d}\big\langle {\bs k}_1, n\ell, \lambda \big| \gamma_{ab,cd}(t) L_{cd}^\dagger L_{ab} \rho_S (0)\big| {\bs k}_2, n\ell, \lambda\big\rangle 
\\
\nn
&=& t \frac{(c_4V_A^s)^2T_F}{M^2N_c}  \int\frac{\diff^3p_{\ma{cm}}}{(2\pi)^3} \frac{\diff^3p_{\ma{rel}}}{(2\pi)^3} \int \diff^4q\,
\delta(E_{n\ell}^\lambda-E_{p_{\rm rel}}-q_0)\delta^3(\bs k_1-{\bs p}_{\rm cm}-{\bs q}) \\
&\times& \tilde{g}^{B++}_{ij}(q_0,{\bs q}) \varepsilon^*_{\lambda i}\varepsilon_{\lambda j}  
\langle \psi^{\lambda}_{n\ell} |  \Psi_{{\bs p}_\ma{rel}} \rangle  \langle \Psi_{{\bs p}_\ma{rel}} | \psi^{\lambda}_{n\ell} \rangle
\big\langle {\bs k}_1, n\ell, \lambda\big| \rho_S(0) \big| {\bs k}_2, n\ell, \lambda \big\rangle \,.
\ee
One can show that $\sum_{a,b,c,d}\big\langle {\bs k}_1, n\ell, \lambda \big|\gamma_{ab,cd}(t) \rho_S (0)L_{cd}^\dagger L_{ab}\big| {\bs k}_2, n\ell, \lambda\big\rangle$ yields the same result. The $\delta$ function for energy conservation does not contain the center-of-mass kinetic energy of the octet unbound state, since it is suppressed as explained below Eq.~\eqref{eqn:ME_B}. Physically it means we neglect the recoil effect~\cite{Biondini:2024aan}.

Now we perform the Wigner transformation and obtain the collision term in the Boltzmann equation 
\begin{align}\label{eq:diss_collisions}
&\ml{C}_{n\ell,\lambda}^-({\bs x}, {\bs k}, t=0)[f_{n\ell, \lambda}]= \lim_{t\rightarrow 0} \int\frac{\diff^3 k'}{(2\pi)^3}e^{i {\bs k}'\cdot {\bs x} } \Big\langle {\bs k}+\frac{{\bs k}'}{2}, n\ell, \lambda \Big| \partial_{t}\tilde{\mathcal{C}}^{-}(t) \Big|{\bs k}-\frac{{\bs k}'}{2}, n\ell, \lambda \Big\rangle \nonumber \\
&= \frac{(c_4V_A^s)^2T_F}{M^2N_c}  \int \frac{\diff^3p_{\ma{rel}}}{(2\pi)^3} \, \tilde{g}^{B++}_{ij}(E_{n\ell}^\lambda-E_p) \varepsilon^*_{\lambda i}\varepsilon_{\lambda j}  
|\langle \psi^{\lambda}_{n\ell} |  \Psi_{{\bs p}_\ma{rel}} \rangle|^2 f_{n\ell,\lambda}({\bs x},{\bs k},t=0) \,. 
\end{align}
The limit $t\to0$ is not contradictory with the upper limit of the above time integral being $t=\infty$ due to the separation between the system relaxation time and the environment correlation time, as explained in Sec.~\ref{sect:2}.
The chromomagnetic field correlator for the inclusive dissociation process shown in Eq.~(\ref{eq:diss_collisions}) is given by
\begin{align}
\tilde{g}^{B++}_{ij}(q_0) &= \int{\rm d}t\,e^{iq_0t} g_{ij}^{B++}(t) \,, \nn\\
g_{ij}^{B++}(t) &= \Tr_E\Big\{ gB_i^a({\bs R},t) W^{ac}[({\bs R},t),({\bs R},\infty)] W^{cb}[({\bs R},\infty),({\bs R},0)] gB_j^b({\bs R},0) \rho_T(0) \Big\} \nn\\
&= \Tr_E\Big\{ gB_i^a({\bs R},t) W^{ab}[({\bs R},t),({\bs R},0)] gB_j^b({\bs R},0) \rho_T(0) \Big\} \,,
\end{align}
where the ${\bs R}$ dependence in the fields is irrelevant due to the spatial translational invariance. As we mentioned earlier, the spatial Wilson line at infinite time disappears, as well as the $t_0$ dependence.
Generalizing the derivation to an arbitrary initial time rather than $t_i=0$, we obtain $\ml{C}_{n\ell,\lambda}^-({\bs x}, {\bs k}, t_i)[f_{n\ell, \lambda}]$ similarly as written above with $\rho(0)$ replaced by $\rho(t_i)$. The environment correlator depends on the initial time through the time dependence of the temperature in the thermal state, as explained in Sec.~\ref{sect:2}.

\subsection{Recombination}
Analogously, the recombination term is given by the $\widetilde{\ml{C}}^+(t)$ term in Eq.~\eqref{eqn:Cpm}. We consider the matrix element $\langle {\bs k}_1, n_1\ell_1, \lambda_1 | L_{ab} \rho_S(0) L_{cd}^\dagger | {\bs k}_2, n_2\ell_2, \lambda_2\rangle $, which immediately yields $|E_a\rangle=|{\bs k}_1, n_1\ell_1, \lambda_1\rangle$ and $| E_c\rangle=|{\bs k}_2, n_2\ell_2, \lambda_2\rangle$. Then, from the structure of $\gamma_{ab,cd}(t)$, we find $|E_d\rangle=|{\bs p}_{2\ma{cm}}, {\bs p}_{2\ma{rel}}, a_2 \rangle$ and $|E_b\rangle=| {\bs p}_{1\ma{cm}}, {\bs p}_{1\ma{rel}}, a_1\rangle$, respectively. More explicitly, we obtain
\begin{align}
\label{eqn:step_r1}\nonumber
& \sum_{a,b,c,d}\big\langle {\bs k}_1, n_1\ell_1, \lambda_1 \big|\gamma_{ab,cd}(t)L_{ab} \rho_S(0) L_{cd}^\dagger\big| {\bs k}_2, n_2\ell_2,\lambda_2\big\rangle
\\\nonumber
&=  \frac{(c_4  V_{A}^s)^2T_F}{M^2N_c} \sum_{a_1,a_2} 
\int\frac{\diff^3p_{1\ma{cm}}}{(2\pi)^3} \frac{\diff^3p_{1\ma{rel}}}{(2\pi)^3} 
\frac{\diff^3p_{2\ma{cm}}}{(2\pi)^3} \frac{\diff^3p_{2\ma{rel}}}{(2\pi)^3} 
\int \diff^3 R_1 \int \diff^3R_2 \int_{0}^{t} \diff t_1 \int_{0}^{t} \diff t_2 \,
\\
&\times   e^{i(E_{n_1\ell_1}^{\lambda_1} t_1 - {\bs k}_1\cdot {\bs R}_1 )  -i(E_{p_{1\rm{rel}}}t_1 - {\bs p}_{1\ma{cm}} \cdot {\bs R}_1) } 
e^{-i(E_{n_2\ell_2}^{\lambda_2}t_2 - {\bs k}_2\cdot {\bs R}_2 )  + i(E_{p_{2\rm{rel}}}t_2 - {\bs p}_{2\ma{cm}} \cdot {\bs R}_2) } \langle \psi^{\lambda_1}_{n_1\ell_1} | \Psi_{{\bs p}_{1\ma{rel}}} \rangle  \nn\\
&\times \langle \Psi_{{\bs p}_{2\ma{rel}}} | \psi^{\lambda_2}_{n_2\ell_2} \rangle 
\big[ g^{B--}_{ji}(t_2, t_1, {\bs R}_2, {\bs R}_1) \big]^{a_2a_1}
\varepsilon^*_{\lambda_1 i}\varepsilon_{\lambda_2j} 
\big\langle {\bs p}_{1\ma{cm}}, {\bs p}_{1\ma{rel}}, a_1 \big| \rho_S(0) \big| {\bs p}_{2\ma{cm}}, {\bs p}_{2\ma{rel}}, a_2 \big\rangle\,, 
\end{align}
where we have flipped $t_1\leftrightarrow t_2$ in the definition of $\gamma_{ab,cd}(t)$ and introduced a new chromomagnetic field correlation function
\begin{align}
\label{eqn:pre_g--}
&\big[ g^{B--}_{ji}(t_2, t_1, {\bs R}_2, {\bs R}_1) \big]^{a_2a_1} \equiv \Tr_E\Big\{ W^{a_2b}[({\bs R}_1,-\infty),({\bs R}_2,-\infty)]  \nn\\
& \times W^{bc}[({\bs R}_2,-\infty),({\bs R}_2,t_2)] gB_j^c({\bs R}_2,t_2) gB_i^d({\bs R}_1,t_1) W^{da_1}[({\bs R}_1,t_1),({\bs R}_1,-\infty)] \rho_T(0) \Big\}  \,,
\end{align}
where the environment density matrix is at the initial time $t_i=0$ and has been assumed to change little during the time step $t$.
We have chosen $t_0\rightarrow-\infty$ since the octet is an initial state, turns to a color-singlet state after recombination, and thus lives in the far past. When the initial state is an octet field in thermal equilibrium with the environment, one should account for the octet charge effect on the environment thermal state, which gives an adjoint Wilson line along the imaginary time at $t=-\infty$, i.e., $W^{ab}(-i\beta-\infty,-\infty)$. This Wilson line is important for deriving the Kubo-Martin-Schwinger relation between $g^{B++}$ and $g^{B--}$. Here we just omit it for simplicity. See the detailed discussions in Ref.~\cite{Scheihing-Hitschfeld:2023tuz}.

Before performing the momentum and spacetime integrals, we make the following simplifications: First, we focus on the diagonal part in $n\ell,\lambda$ since we want the distribution function $f_{n\ell,\lambda}$ by applying a Wigner transformation, so we set $n_1=n_2=n$, $\ell_1=\ell_2=\ell$, and $\lambda_1=\lambda_2=\lambda$. Second, we assume the system density matrix is diagonal in the color space
\begin{align}
\label{eqn:color_semi}
\big\langle {\bs p}_{1\ma{cm}}, {\bs p}_{1\ma{rel}}, a_1 \big| \rho_S(0) \big| {\bs p}_{2\ma{cm}}, {\bs p}_{2\ma{rel}}, a_2 \big\rangle
\approx \delta^{a_1a_2}
\big\langle {\bs p}_{1\ma{cm}}, {\bs p}_{1\ma{rel}} \big| \rho^{(8)}_S(0) \big| {\bs p}_{2\ma{cm}}, {\bs p}_{2\ma{rel}} \big\rangle \,,
\end{align}
where the superscript $(8)$ indicates that the density matrix describes a color-octet state. Finally, we assume the environment correlator is translationally invariant in spacetime and then perform a Fourier transform
\begin{align}
\tilde{g}^{B--}_{ji}(q)&=\int \diff\delta t \diff^3\delta R \, e^{-iq_0\delta t+i\bs q\cdot\delta\bs R} g^{B--}_{ji}(t_2, t_1, {\bs R}_2, {\bs R}_1)\,,\nn\\
\label{eqn:gB--}
g^{B--}_{ji}(t_2, t_1, {\bs R}_2, {\bs R}_1) &\equiv \big[g^{B--}_{ji}(t_2, t_1, {\bs R}_2, {\bs R}_1)\big]^{aa} \,,
\end{align}
where the superscript color index $a$ is summed over, $\delta t = t_1-t_2$, $\delta{\bs R} = {\bs R}_1-{\bs R}_2$ as defined for the $g^{B++}$ case, and we have used the assumption that the octet density matrix is diagonal in the color space.

Following the same approach for handling the dissociation term, we acquire
\begin{align}
\label{eqn:step_r2}
& \sum_{a,b,c,d}\big\langle {\bs k}_1, n\ell, \lambda \big|\gamma_{ab,cd}(t)L_{ab} \rho_S(0) L_{cd}^\dagger\big| {\bs k}_2, n\ell, \lambda\big\rangle \nn\\
&= \frac{(c_4  V_{A}^s)^2 T_F}{M^2N_c}
\int \frac{\diff^3p_{1\ma{rel}}}{(2\pi)^3} 
\frac{\diff^3p_{2\ma{rel}}}{(2\pi)^3} 
\int \frac{\diff^4q}{(2\pi)^4} (2\pi)^2\delta(E_{n\ell}^\lambda-E_{p_{1\rm{rel}}}+q_0) \delta(E_{n\ell}^\lambda-E_{p_{2\rm{rel}}}+q_0) \nn\\
&\times   
\langle \psi^{\lambda}_{n\ell} | \Psi_{{\bs p}_{1\ma{rel}}} \rangle  
\langle \Psi_{{\bs p}_{2\ma{rel}}} | \psi^{\lambda}_{n\ell} \rangle \tilde{g}^{B--}_{ji}(q) \varepsilon_{\lambda i}^*\varepsilon_{\lambda j}
\big\langle {\bs k}_1+{\bs q}, {\bs p}_{1\ma{rel}} \big| \rho_S^{(8)}(0) \big| {\bs k}_2+{\bs q}, {\bs p}_{2\ma{rel}} \big\rangle\,.
\end{align}
Taking the Wigner transformation gives the recombination term in Eq.~(\ref{eqn:Boltzmann_p}),
\begin{align}
&\ml{C}_{n\ell,\lambda}^+({\bs x}, {\bs k}, t=0)[f^{(8)}]=\lim_{t\rightarrow 0}\int\frac{\diff^3 k'}{(2\pi)^3}e^{i {\bs k}'\cdot {\bs x} } \Big\langle {\bs k}+\frac{{\bs k}'}{2}, n\ell, \lambda\Big|\partial_{t}\widetilde{\mathcal{C}}^{+}(t)\Big|{\bs k}-\frac{{\bs k}'}{2}, n\ell, \lambda\Big\rangle \,.
\end{align}
We now introduce the distribution function for the color-octet $Q\bar{Q}$ pairs,
\be
\label{eqn:wigner8}
&&\int\frac{\diff^3 k'}{(2\pi)^3} e^{i{\bs k}'\cdot{\bs x}} \Big\langle {\bs k}+{\bs q}+\frac{{\bs k}'}{2}, {\bs p}_{1\ma{rel}} \Big| \rho^{(8)}_S(t) \Big| {\bs k}+{\bs q}-\frac{{\bs k}'}{2}, {\bs p}_{2\ma{rel}} \Big\rangle \nn\\
&=& \int \diff^3 x_\ma{rel} e^{-i({\bs p}_{1\ma{rel}} - {\bs p}_{2\ma{rel}}) \cdot {\bs x}_\ma{rel} } f_{Q\bar{Q}}^{(8)} \Big({\bs x}, {\bs k}+{\bs q}, {\bs x}_\ma{rel}, \frac{ {\bs p}_{1\ma{rel}}+{\bs p}_{2\ma{rel}} }{2}, t \Big) \,,
\ee
where ${\bs x}$ and ${\bs k}+{\bs q}$ are the center-of-mass position and momentum, respectively.
We then adopt the gradient expansion by assuming the distribution varies slowly as ${\bs x}_{\ma{rel}}$ changes and expanding $f_{Q\bar{Q}}^{(8)}$ around some ${\bs x}_{\ma{rel}} = {\bs x}_0$ (usually chosen to be ${\bs x}_0={\bs 0}$),
\be
\label{eqn:gradient}
f_{Q\bar{Q}}^{(8)}({\bs x}, {\bs p}_{\rm cm}, {\bs x}_\ma{rel}, \frac{ {\bs p}_{1\ma{rel}}+{\bs p}_{2\ma{rel}} }{2}, t ) 
= f_{Q\bar{Q}}^{(8)}({\bs x}, {\bs p}_{\rm cm}, {\bs x}_0, \frac{ {\bs p}_{1\ma{rel}}+{\bs p}_{2\ma{rel}} }{2} ,t ) \nn\\
+ ( {\bs x}_\ma{rel} - {\bs x}_0) \cdot \nabla_{{\bs x}_0} f_{Q\bar{Q}}^{(8)}({\bs x}, {\bs p}_{\rm cm}, {\bs x}_0, \frac{ {\bs p}_{1\ma{rel}}+{\bs p}_{2\ma{rel}} }{2},t ) + \cdots \,,
\ee
where higher-order terms in the gradient expansion are omitted. Keeping only the leading-order distribution, we have 
\be
&&\int\frac{\diff^3 k'}{(2\pi)^3} e^{i{\bs k}'\cdot{\bs x}} \Big\langle {\bs k}+{\bs q}+\frac{{\bs k}'}{2}, {\bs p}_{1\ma{rel}} \Big| \rho^{(8)}_S(t) \Big| {\bs k}+{\bs q}-\frac{{\bs k}'}{2}, {\bs p}_{2\ma{rel}} \Big\rangle \nn\\
&\approx& (2\pi)^3 \delta^3({\bs p}_{1\ma{rel}} - {\bs p}_{2\ma{rel}}) f_{Q\bar{Q}}^{(8)}({\bs x}, {\bs k}+{\bs q}, {\bs x}_0, \frac{ {\bs p}_{1\ma{rel}}+{\bs p}_{2\ma{rel}} }{2} ,t ) \,.
\ee
Because of the $\delta$ function for the relative momenta, we have ${\bs p}_{1\ma{rel}} ={\bs p}_{2\ma{rel}}={\bs p}_{\ma{rel}} $ and thus $E_{p_{1\rm{rel}}}=E_{p_{2\rm{rel}}}=E_{p_{\rm rel}}$.
At LO in the gradient expansion, we eventually arrive at
\begin{align}\nonumber
\ml{C}_{n\ell,\lambda}^+({\bs x}, {\bs k}, t=0)[f^{(8)}]& = \frac{(c_4 V_{A}^s)^2T_F}{M^2N_c} \!\!\int\frac{\diff^3p_{\ma{cm}}}{(2\pi)^3} \frac{\diff^3p_{\ma{rel}}}{(2\pi)^3} \!
\int\! \diff^4q \, \delta(E_{n\ell}^\lambda-E_{p_{\rm{rel}}}+q_0)\delta^3({\bs k}-{\bs p}_{\rm cm}+{\bs q}) \nn\\
&\times |\langle \psi^{\lambda}_{n\ell} | \Psi_{{\bs p}_{\ma{rel}}} \rangle|^2 
\tilde{g}^{B--}_{ji}(q) \varepsilon_{\lambda i}^* \varepsilon_{\lambda j} f_{Q\bar{Q}}^{(8)}({\bs x}, {\bs p}_{\ma{cm}}, {\bs x}_0, {\bs p}_{\ma{rel}}, t = 0) \,.
\end{align}
In practical applications, one can choose ${\bs x}_{0}={\bs 0}$. Note that here $\tilde{g}^{B--}_{ji}(q)$ is the sum of the diagonal components in color space as defined in Eq.~\eqref{eqn:gB--}. Integrating over ${\bs p}_{\ma{cm}}$, we arrive at
\begin{align}
\label{eq:recomb_collisions}
\ml{C}_{n\ell,\lambda}^+({\bs x}, {\bs k}, t=0)[f^{(8)}] 
&= \frac{(c_4V_A^s)^2T_F}{M^2N_c}  \int \frac{\diff^3p_{\ma{rel}}}{(2\pi)^3} \int \diff^3 q \,  \tilde{g}^{B--}_{ji}(q_0=E_{p_{\rm rel}}-E_{n\ell}^\lambda,\,{\bs q})  \varepsilon^*_{\lambda i}\varepsilon_{\lambda j}  
\nn
\\
&\times |\langle \psi^{\lambda}_{n\ell} |  \Psi_{{\bs p}_\ma{rel}} \rangle|^2 
f_{Q\bar{Q}}^{(8)}({\bs x}, {\bs k+\bs q}, {\bs x}_0, {\bs p}_{\ma{rel}}, t = 0)\,. 
\end{align}
However, due to the $\bs q$ dependence of $f_{Q\bar{Q}}^{(8)}$, we are unable to write $\ml{C}_{n\ell,\lambda}^+({\bs x}, {\bs k}, t=0)[f^{(8)}]$ as the simpler form for $\ml{C}_{n\ell,\lambda}^-({\bs x}, {\bs k}, t=0)[f_{nl,\lambda}]$.
The generalization to an arbitrary initial time instead of $t_i=0$ is straightforward, which gives $\ml{C}_{n\ell,\lambda}^+({\bs x}, {\bs k}, t)[f^{(8)}]$ in a similar fashion.

\subsection{Full collision term}
Generically, we may also incorporate the spin-independent collision terms generated by chromoelectric fields. 
To this end, one simply needs to append the spin indices to the result in Ref.~\cite{Yao:2020eqy} for the transition between the color-singlet vector quarkonium states and color-octet unbound pairs. As will we see, the transition probability induced by the chromoelectric interaction scales as $\frac{r^2}{M^0}$ in the power counting while that induced by the chromomagnetic interaction scales as $\frac{r^0}{M^2}$.

The full dissociation term accordingly can be written as
\begin{align}
\label{eq:C_dissociation}
&\ml{C}_{n\ell,\lambda}^-({\bs x}, {\bs k}, t)[f_{n\ell,\lambda}] \nn\\
& = \int\frac{\diff^3p_{\ma{cm}}}{(2\pi)^3} \frac{\diff^3p_{\ma{rel}}}{(2\pi)^3}  \int \diff^4q \, \delta(E_{n\ell}^\lambda-E_{p_{\rm rel}}-q_0)
\delta^3({\bs k}-{\bs p}_{\rm cm}-{\bs q})
|\ml{M}_d|^2 f_{n\ell,\lambda}({\bs x}, {\bs k}, t) \,,
\end{align} 
where the transition amplitudes squared are defined as
\begin{align}
&|\ml{M}_d|^2 \nn\\
&= \frac{V_A^2T_F}{N_c}\tilde{g}^{E++}_{ij}(q)\langle \psi^{\lambda}_{n\ell} | r_i | \Psi_{{\bs p}_\ma{rel}}^{\lambda} \rangle 
\langle \Psi_{{\bs p}_\ma{rel}}^{\lambda} | r_j | \psi^{\lambda}_{n\ell} \rangle + \frac{(c_4  V_{A}^s)^2T_F}{M^2N_c}\tilde{g}^{B++}_{ij}(q) \varepsilon^*_{\lambda i}\varepsilon_{\lambda j}  
|\langle \psi^{\lambda}_{n\ell} |  \Psi_{{\bs p}_\ma{rel}} \rangle|^2 \,,
\end{align}
where $\tilde{g}^{E++}_{ij}(q)$ denotes the chromoelectric field correlator, analogous to $\tilde{g}^{B++}_{ij}(q)$ with the chromomagnetic fields replaced by chromoelectric fields. In the electric process, the dissociated $Q\bar{Q}$ still carries the spin polarization $\lambda$, while in the magnetic process, the dissociated $Q\bar{Q}$ becomes a spin singlet. This difference is reflected in the superscript $\lambda$ for the unbound wave function $\Psi_{{\bs p}_{\rm rel}}$.

Similarly, the recombination term reads
\begin{align}
\label{eq:C_recombination}
&\ml{C}_{n\ell,\lambda}^+({\bs x}, {\bs k}, t)[f^{(8)},f^{(8)}_\lambda] = \int\frac{\diff^3p_{\ma{cm}}}{(2\pi)^3} \frac{\diff^3p_{\ma{rel}}}{(2\pi)^3} \int \diff^4q \,  \delta(E_{n\ell}^\lambda -E_{p_{\rm rel}}+q_0)\delta^3({\bs k}-{\bs p}_{\rm cm}+{\bs q}) \nn\\
&\qquad\qquad \times \left( |\ml{M}_{r,e}|^2 f_{Q\bar{Q}\lambda}^{(8)}({\bs x}, {\bs p}_{\ma{cm}}, {\bs x}_0, {\bs p}_{\ma{rel}} ,t) + |\ml{M}_{r,b}|^2 f_{Q\bar{Q}}^{(8)}({\bs x}, {\bs p}_{\ma{cm}}, {\bs x}_0, {\bs p}_{\ma{rel}} ,t) \right) \,,
\end{align}
where
\begin{align}
|\ml{M}_{r,e}|^2 &= \frac{V_A^2T_F}{N_c} \tilde{g}^{E--}_{ji}(q)\langle \psi^{\lambda}_{n\ell} | r_i | \Psi_{{\bs p}_\ma{rel}}^{\lambda} \rangle \langle \Psi_{{\bs p}_\ma{rel}}^{\lambda} | r_j | \psi^{\lambda}_{n\ell} \rangle \,, \nn\\
|\ml{M}_{r,b}|^2 &= \frac{(c_4  V_{A}^s)^2T_F}{M^2N_c}\tilde{g}^{B--}_{ji}(q) \varepsilon^*_{\lambda i}\varepsilon_{\lambda j} |\langle \psi^{\lambda}_{n\ell} |  \Psi_{{\bs p}_\ma{rel}} \rangle|^2 \,.
\end{align}
The chromoelectric correlator $\tilde{g}^{E--}_{ij}(q)$ is similar to $\tilde{g}^{B--}_{ij}(q)$, defined by replacing the chromomagnetic fields with chromoelectric fields. Here $f_{Q\bar{Q}}^{(8)}$ represents the distribution function for the color-octet spin-singlet $Q\bar{Q}$ pairs while $f_{Q\bar{Q}\lambda}^{(8)}$ is for the color-octet spin-triplet ones. 
Even though the chromoelectric fields do not couple to the polarization vectors for both the dissociation and recombination terms as opposed to the chromomagnetic fields that explicitly modify the polarization transport, the electric interaction may still implicitly affect the polarization especially in the presence of a nontrivial $f_{Q\bar{Q}\lambda}^{(8)}$ originating from initial polarization of the $Q\bar{Q}$ pair or far-from-equilibrium dynamics in the initial stage of heavy ion collisions.

\section{Lindblad equation in the quantum Brownian motion limit}
\label{sect:5}
In this section, we work under the energy scale hierarchy $M\gg Mv \gg Mv^2,T,\Lambda_{\rm QCD}$ and further assume $T\gg Mv^2$, which justifies an expansion in $\frac{Mv^2}{T}$ and the Quantum Brownian motion limit. 
Most previous studies of quarkonium in the quantum Brownian motion limit integrated out the center-of-mass motion, which we will follow here. By making the relative position dependence implicit, we can write the Hamiltonian as
\begin{align}
\label{eqn:HS_qbm}
H_S & = \frac{{\bs p}^2_{\rm rel}}{M} + V_s^1(r) |s \rangle\langle s| +  V_s^\lambda(r) |s,\lambda \rangle\langle s,\lambda| + V_o^1(r) |a \rangle\langle a| + V_o^\lambda(r) |a,\lambda \rangle\langle a,\lambda| \,, \nn\\
\widetilde{H}_I(t) &= \frac{c_4V_A^s}{M} \sqrt{\frac{T_F}{N_c}} \left( g\widetilde{B}_i^a(t) \big( \varepsilon_{\lambda i}|s\rangle\langle a,\lambda| + \varepsilon_{\lambda i}^*|s,\lambda\rangle \langle a|\big) + g\widetilde{B}_i^{a\dagger}(t) \big( \varepsilon_{\lambda i}^*|a,\lambda\rangle \langle s| + \varepsilon_{\lambda i}|a\rangle \langle s,\lambda| \big) \right)  \nn\\
&+ \frac{c_4V_B^s}{2M} d^{abc} g(WB_iW)^{a'abcc'}(t) \big( \varepsilon_{\lambda i}|a'\rangle \langle c',\lambda| + \varepsilon_{\lambda i}^* |a',\lambda\rangle \langle c'| \big) \nn\\
& - \frac{c_4V_B^s}{2M}f^{abc}\varepsilon_{ijk} g(WB_jW)^{a'abcc'}(t) \varepsilon_{\lambda i}^* \varepsilon_{\lambda'k} |a',\lambda\rangle \langle c',\lambda'| \,,
\end{align}
where $i$, $\lambda$, and $\lambda'$ are implicitly summed over. The tilted magnetic fields and $WB_iW$ are defined by 
\begin{align}
&\widetilde{B}^a_i(t) = B^b_i(t) W^{ba}(t, -\infty) \,,\qquad \widetilde{B}^{a\dagger}_i(t) = W^{ab}(\infty, t) B^b_i(t) \,, \nn\\
& (WB_iW)^{a'abcc'}(t) = W^{a'a}(\infty, t) B^b_i(t) W^{cc'}(t, -\infty) \,,
\end{align}
where we have omitted the center-of-mass position dependence.
The end point of the adjoint timelike Wilson line is either $t_0=-\infty$ when the initial state is an octet state or $t_0=\infty$ when the final state is an octet state, as explained in the previous section.
The structure of $WB_iW$ is similar to the $UE_iU$ structure in the definition of the heavy quark diffusion coefficient where $U$ denotes a fundamental Wilson line along a straight timeline. The similarity originates from that in the octet-octet transition process, both the initial and final states carry colors, as in the heavy quark diffusion process. As a result, the operator ordering for diffusion is different from that for dissociation and recombination, which crucially gives different environment correlators with different values even at NLO in coupling~\cite{Eller:2019spw,Scheihing-Hitschfeld:2022xqx}.

From the Hamiltonian expressions, we can easily find out the system and environment operators. Plugging them into the general Lindblad equation shown in Sec.~\ref{sect:qbm} gives the Lindblad equation for the quarkonium polarization. In a simple scenario where the density matrix is diagonal in both the color and spin spaces, we can greatly simplify the Lindblad equation. More specifically, we assume
\begin{multline}
\label{eqn:rho_diag_qbm}
\rho_S(t) = \left(
  \begin{matrix}
\langle s| \rho_S |s\rangle & 0 & 0 & 0 & 0 \\
0 & \langle s,+| \rho_S |s,+\rangle & 0 & 0 & 0 \\
0 & 0 & \langle s,-| \rho_S |s,-\rangle & 0 & 0 \\
0 & 0 & 0 & \langle s,0| \rho_S |s,0\rangle & 0 \\
0 & 0 & 0 & 0 & \sum_a\langle a | \rho_S |a\rangle \\
0 & 0 & 0 & 0 & 0 \\
0 & 0 & 0 & 0 & 0 \\
0 & 0 & 0 & 0 & 0
  \end{matrix}\right.\\
\left.\begin{matrix}
0 & 0 & 0\\
0 & 0 & 0\\
0 & 0 & 0\\
0 & 0 & 0\\
0 & 0 & 0\\
\sum_a \langle a,+| \rho_S |a,+\rangle & 0 & 0 \\
0 & \sum_a \langle a,-| \rho_S |a,-\rangle & 0 \\
0 & 0 & \sum_a \langle a,0| \rho_S |a,0\rangle
  \end{matrix}\right) \,.
\end{multline}
Then the Lindblad equation at LO in the quantum Brownian motion expansion can be simplified as
\begin{align}
\frac{{\rm d}\rho_S(t)}{{\rm d}t} = -i [H_S+\Delta H_S, \rho_S(t)] + K_{\alpha} \left( L_{\alpha} \rho_S(t) L_{\alpha}^\dagger -\frac{1}{2} \{L_{\alpha}^\dagger L_{\alpha}, \rho_S(t) \} \right) \,,
\end{align}
where $\Delta H_S$, $K_{\alpha}$, and $L_{\alpha}$ can be obtained from $\widetilde{H}_I(t)$ in Eq.~(\ref{eqn:HS_qbm}) by using relevant equations in Sec.~\ref{sect:qbm}. More specifically, the Hamiltonian in the Lindblad equation is given by
\begin{align}
H_S = -\frac{{\bs \nabla}_{\rm rel}^2}{M}I + \begin{pmatrix}
V_s^1(r) & 0 & 0 & 0 & 0 & 0 & 0 & 0\\
0 & V_s^+(r) & 0 & 0 & 0 & 0 & 0 & 0\\
0 & 0 & V_s^-(r) & 0 & 0 & 0 & 0 & 0 \\
0 & 0 & 0 & V_s^0(r) & 0 & 0 & 0 & 0 \\
0 & 0 & 0 & 0 & V_o^1(r) & 0 & 0 & 0 \\
0 & 0 & 0 & 0 & 0 & V_o^+(r) & 0 & 0\\
0 & 0 & 0 & 0 & 0 & 0 & V_o^-(r) & 0\\
0 & 0 & 0 & 0 & 0 & 0 & 0 & V_o^0(r) \\
\end{pmatrix} \,,
\end{align}
where $I$ denotes the identity matrix.
The correction to the Hamiltonian is given by
\begin{align}
\Delta H_S &= {\rm diag}(\Delta H_{s}^1, \Delta H_{s}^{+}, \Delta H_{s}^{-}, \Delta H_{s}^{0},\Delta H_{o}^1, \Delta H_{o}^+, \Delta H_{o}^-, \Delta H_{o}^0) \,, \nn\\
\Delta H_{s}^1 &= \frac{(c_4V_A^s)^2T_F}{2M^2N_c} \gamma_{B,ii}^{++} \,, \qquad\qquad\quad \Delta H_{s}^\lambda = \frac{(c_4V_A^s)^2T_F}{2M^2N_c} \gamma_{B,ij}^{++} \varepsilon_{\lambda i}^*\varepsilon_{\lambda j} \,, \nn\\
\Delta H_{o}^1 &= \frac{(c_4V_A^s)^2T_F}{2M^2N_c} \frac{\gamma_{B,ii}^{--}}{N_c^2-1} + \frac{(c_4V_B^s)^2}{8M^2} \frac{\gamma_{Bd,ii}^{-++-}}{N_c^2-1} \,, \nn\\
\Delta H_{o}^\lambda &= \left( \frac{(c_4V_A^s)^2T_F}{2M^2N_c} \frac{\gamma_{B,ij}^{--}}{N_c^2-1} + \frac{(c_4V_B^s)^2}{8M^2} \frac{\gamma_{Bd,ij}^{-++-}}{N_c^2-1} \right) \varepsilon_{\lambda i}^*\varepsilon_{\lambda j} + \frac{(c_4V_B^s)^2}{8M^2} \frac{\gamma_{Bf,ji}^{-++-}}{N_c^2-1}( \varepsilon_{\lambda i}^*\varepsilon_{\lambda j}-\delta_{ij}) \,,
\end{align}
where $\lambda$ is not summed, and we have used $\sum_{\lambda'}\varepsilon^*_{\lambda' i}\varepsilon_{\lambda' j}=\delta_{ij}$ when $\lambda'$ is summed over for intermediate states. The $\frac{1}{N_c^2-1}$ color average factor is included in the octet channel to avoid double counting: Color indices are summed both in the chromomagnetic correlators and in the density matrix as shown in Eq.~\eqref{eqn:rho_diag_qbm}. The environment correlation functions (transport coefficients $\gamma_B$) are defined as\footnote{As explained earlier, we omit the imaginary time adjoint Wilson lines at $t=-\infty$, as will be similarly done for the $\kappa_B$ coefficients to be defined in Eq.~\eqref{eqn:kappa_B}. One can think of them as part of $\rho_T$ when the initial state is a color octet. An explicit expression with the imaginary time Wilson line included can be found in Ref.~\cite{Scheihing-Hitschfeld:2023tuz}.}
\begin{align}
\gamma_{B,ij}^{++} &= -ig^2\int_{-\infty}^{\infty} {\rm d}t \, {\rm sign}(t) {\rm Tr}\left( B_i^{a}(t) W^{ab}(t,+\infty) W^{bc}(+\infty,0) B_j^c(0) \rho_T \right) \,, \nn\\
\gamma_{B,ij}^{--} &= -ig^2\int_{-\infty}^{\infty} {\rm d}t \, {\rm sign}(t) {\rm Tr}\left( W^{ab}(-\infty,t) B_i^{b}(t) B_j^c(0) W^{ca}(0,-\infty)  \rho_T \right) \,, \nn\\
\gamma_{Bd,ij}^{-++-} &= -ig^2 d^{abc}d^{def} \int_{-\infty}^{\infty} {\rm d}t \, {\rm sign}(t) {\rm Tr}\left( W^{a'a}(-\infty,t)B_i^b(t)W^{cc'}(t,+\infty) \right.  \nn\\
& \qquad\qquad\qquad\qquad\qquad\qquad\qquad \left. W^{c'd}(+\infty,0)B_j^e(0)W^{fa'}(0,-\infty) \rho_T\right) \,,  \nn\\
\gamma_{Bf,ij}^{-++-} &= -ig^2 f^{abc}f^{def} \int_{-\infty}^{\infty} {\rm d}t \, {\rm sign}(t) {\rm Tr}\left( W^{a'a}(-\infty,t)B_i^b(t)W^{cc'}(t,+\infty) \right. \nn\\
& \qquad\qquad\qquad\qquad\qquad\qquad\qquad \left. W^{c'd}(+\infty,0)B_j^e(0)W^{fa'}(0,-\infty) \rho_T\right) \,.
\end{align}
The relevant Lindblad operators are
\begin{align}
(L_{1+}, L_{1-}, L_{10}) &= \frac{c_4V_A^s}{M}\sqrt{\frac{T_F}{N_c}} \frac{1}{\sqrt{N_c^2-1}} (T_{16}, T_{17}, T_{18}) \,, \nn\\
(L_{2+}, L_{2-}, L_{20}) &= \frac{c_4V_A^s}{M}\sqrt{\frac{T_F}{N_c}} (T_{61}, T_{71}, T_{81}) \,,  \nn\\
(L_{3+}, L_{3-}, L_{30}) &= \frac{c_4V_A^s}{M}\sqrt{\frac{T_F}{N_c}} \frac{1}{\sqrt{N_c^2-1}} (T_{25}, T_{35}, T_{45}) \,, \nn\\
(L_{4+}, L_{4-}, L_{40}) &= \frac{c_4V_A^s}{M}\sqrt{\frac{T_F}{N_c}} (T_{52}, T_{53}, T_{54}) \,, \nn\\
(L_{5+}, L_{5-}, L_{50}) &= \frac{c_4V_B^s}{2M} \frac{1}{\sqrt{N_c^2-1}} (T_{56}, T_{57}, T_{58}) \,, \nn\\
(L_{6+}, L_{6-}, L_{60}) &= \frac{c_4V_B^s}{2M} \frac{1}{\sqrt{N_c^2-1}} (T_{65}, T_{75}, T_{85}) \,, \nn\\
(L_{7++}, L_{7+-}, L_{7+0}) &= \frac{c_4V_B^s}{2M} \frac{1}{\sqrt{N_c^2-1}} (T_{66},   T_{67}, T_{68}) \,, \nn\\
(L_{7-+}, L_{7--}, L_{7-0}) &= \frac{c_4V_B^s}{2M} \frac{1}{\sqrt{N_c^2-1}} (T_{76},   T_{77}, T_{78}) \,, \nn\\
(L_{70+}, L_{70-}, L_{700}) &= \frac{c_4V_B^s}{2M} \frac{1}{\sqrt{N_c^2-1}} (T_{86},   T_{87}, T_{88}) \,,
\end{align}
where the $\frac{1}{\sqrt{N_c^2-1}}$ factor is included to avoid double counting for an initial octet state as explained above, and $T_{ij}$ denotes a $8\times8$ matrix whose entries are zeros except at the $i$th row and $j$th column where the entry is $1$. The $L_{1\lambda}$, $L_{2\lambda}$, $L_{3\lambda}$, and $L_{4\lambda}$ operators correspond to the first line of $\widetilde{H}_I(t)$ in Eq.~\eqref{eqn:HS_qbm}, the $L_{5\lambda}$ and $L_{6\lambda}$ operators correspond to the second line of $\widetilde{H}_I(t)$, and the $L_{7\lambda}$ operators correspond to the last line of $\widetilde{H}_I(t)$.
The corresponding environment correlation functions are
\begin{align}
&K_{1\lambda} = \kappa^{--}_{B,ij} \varepsilon^*_{\lambda i} \varepsilon_{\lambda j} \,,& 
&K_{2\lambda} = \kappa^{++}_{B,ij} \varepsilon_{\lambda i} \varepsilon^*_{\lambda j} \,,& 
&K_{3\lambda} = \kappa^{--}_{B,ij} \varepsilon_{\lambda i} \varepsilon^*_{\lambda j} \,,& \nn\\
&K_{4\lambda} = \kappa^{++}_{B,ij} \varepsilon^*_{\lambda i} \varepsilon_{\lambda j} \,,& 
&K_{5\lambda} = \kappa_{Bd,ij}^{-++-} \varepsilon^*_{\lambda i} \varepsilon_{\lambda j} \,,& 
&K_{6\lambda} = \kappa_{Bd,ij}^{-++-} \varepsilon_{\lambda i} \varepsilon^*_{\lambda j} \,,& \nn\\
&K_{7\lambda\lambda'} = \kappa_{Bf,jj'}^{-++-} \varepsilon_{ijk}&& \kern-2.5em \varepsilon_{i'j'k'} \varepsilon_{\lambda' i}^* \varepsilon_{\lambda k} \varepsilon_{\lambda i'}^*  \varepsilon_{\lambda' k'} \,,&&&&
\end{align}
where the transport coefficients are defined as
\begin{align}
\label{eqn:kappa_B}
&\kappa^{--}_{B,ij} = g^2\int_{-\infty}^{\infty} {\rm d}t \, {\rm Tr}\left( W^{ab}(-\infty,t) B_i^{b}(t) B_j^c(0) W^{ca}(0,-\infty)  \rho_T \right) \,, \nn\\
&\kappa^{++}_{B,ij} = g^2\int_{-\infty}^{\infty} {\rm d}t \, {\rm Tr}\left( B_i^{a}(t) W^{ab}(t,+\infty) W^{bc}(+\infty,0) B_j^c(0) \rho_T \right) \,, \nn\\
&\kappa_{Bd,ij}^{-++-} =\nn\\
& g^2 d^{abc}d^{def} \int_{-\infty}^{\infty} {\rm d}t \, {\rm Tr}\left( W^{a'a}(-\infty,t)B_i^b(t)W^{cc'}(t,+\infty) W^{c'd}(+\infty,0)B_j^e(0)W^{fa'}(0,-\infty) \rho_T\right) \,, \nn\\
&\kappa_{Bf,ij}^{-++-} =\nn\\
& g^2 f^{abc}f^{def} \int_{-\infty}^{\infty} {\rm d}t \, {\rm Tr}\left( W^{a'a}(-\infty,t)B_i^b(t)W^{cc'}(t,+\infty) W^{c'd}(+\infty,0)B_j^e(0)W^{fa'}(0,-\infty) \rho_T\right)\,.
\end{align}

The Lindblad equation at NLO in the quantum Brownian motion expansion can be written down explicitly by using Eq.~\eqref{eqn:Lindblad_NLO}. The new contribution is the commutator $[H_S, L_\alpha]$ term which can be easily worked out by using the above formulas. Writing them out explicitly is tedious but straightforward so we will not do so here. We note that when the potentials are spin independent, all the commutator terms vanish for the magnetic interaction. Since the difference between the potentials in the two spin channels is suppressed as $V^1-V^\lambda \sim \frac{1}{M^2}$ for the $S$-wave quarkonium states, we expect the NLO effect in the quantum Brownian motion expansion is very small, suppressed by $\frac{1}{M^3}$.

\section{Conclusions and outlook}
\label{sect:summary}
In this paper, we derive the semiclassical Boltzmann equation in the quantum optical limit ($M\gg Mv \gg Mv^2,T,\Lambda_{\rm QCD}$) and the Lindblad equation in the quantum Brownian limit ($M\gg Mv \gg Mv^2,T,\Lambda_{\rm QCD}$ and $T\gg Mv^2$) for spin-singlet and spin-triplet (possibly polarized) quarkonium states in the QGP by using the open quantum system framework and pNRQCD. 
For simplicity and practical applications to the spin alignment phenomena, we only consider the diagonal components in the spin density matrix of the quarkonium state. In the derived semiclassical Boltzmann equation that tracks the evolution of the polarized distribution function of a $S$-wave vector quarkonium state, the chromomagnetic interaction in pNRQCD gives rise to the dissociation and recombination terms describing the transition between a color-singlet spin-triplet quarkonium state and a color-octet spin-singlet unbound $Q\bar{Q}$ pair. The dissociation and recombination terms depend on the chromomagnetic correlators, which are analogous to the chromoelectric ones. They can also be thought of as generalized gluon distribution functions of the thermal QGP~\cite{Nijs:2023dbc}. Similar interaction terms containing the chromomagnetic correlators are also derived for the Lindblad equation in the quantum Brownian motion limit. Only the zero frequency part of the chromomagnetic correlators appears in the Lindblad equation, which serves as new transport coefficients for heavy flavors.

Future work should calculate the chromomagnetic correlators in both weakly coupled and strongly coupled QGP, as have been done for the chromoelectric correlators by finite temperature perturbation theory~\cite{Binder:2021otw} and AdS/CFT correspondence~\cite{Nijs:2023dks,Nijs:2023dbc}. Some preliminary results via lattice QCD methods can be found in Ref.~\cite{Leino:2024pen}. (The lattice setup was discussed in Ref.~\cite{Scheihing-Hitschfeld:2023tuz}. It is worth mentioning that Refs.~\cite{Mayer-Steudte:2021hei,Altenkort:2024spl} studied a different chromomagnetic correlator, which describes the relativistic correction to the heavy quark diffusion coefficient. The correlators for quarkonia and heavy quarks differ in terms of operator orderings in the Wilson lines~\cite{Scheihing-Hitschfeld:2022xqx}.)

For practical applications to the spin alignment of quarkonia, we shall recall that one of the essential ingredients for the underlying mechanism is the anisotropic effect. In general, the chromomagnetic or chromoelectric fields in the QGP are isotropic in the laboratory frame unless we further take the medium expansion into account. Nevertheless, in the rest frame of quarkonium, such background fields may become anisotropic due to the Lorentz boost that depends on the quarkonium momentum, even though they are isotropic in the laboratory frame. Consequently, when solving the semiclassical Boltzmann equation in the rest frame for a vector quarkonium with finite laboratory-frame momentum, the anisotropic chromomagnetic fields could yield nontrivial $\lambda$ dependence of the collision terms in connection to spin alignment. In fact, the spin alignment signature is measured in the rest frame of the vector meson. Boosting the background fields involved in the quark coalescence scenario for spin alignment was adopted in e.g., Refs.~\cite{Sheng:2022wsy,Kumar:2023ghs}. For quarkonia with small momenta in the laboratory frame, the $\lambda$ dependence of the collision terms becomes trivial, where all spin-triplet states have almost the same collision terms. Under this circumstance, the spin alignment could only originate from the intrinsically anisotropic gluon fields such as the turbulent fields induced by anisotropic expansion of the QGP or the anisotropic glasma fields in the initial stage of heavy ion collisions. On the other hand, as briefly mentioned in Sec.~\ref{sect:Boltzmann_eq}, the electric interaction does not explicitly lead to $\lambda$-dependent collision terms since the chromoelectric fields are not coupled to the polarization vectors. Nonetheless, the corresponding recombination term depends on the color-octet distribution function with polarization dependence, which is in connection to the initial polarization of the heavy quark and antiquark pair. Considering e.g., the glasma effect upon the heavy quark and antiquark pair, the recombination term associated with the chromoelectric interaction may also result in spin alignment, analogous to the simplified coalescence model proposed in Ref.~\cite{Kumar:2023ghs}. Overall, the constructed formalism in this paper will be useful for further investigations on the polarization phenomena of heavy quarkonia especially for the spin alignment. Future applications to the heavy ion phenomenology will be studied.

\acknowledgments
D.-L. Y.~thanks the participants of ``ExHIC-p workshop on polarization phenomena in nuclear collisions'' in 2024 for fruitful discussions, especially Y. Akamatsu and S. Lin for sharing their recent studies on spin transport of heavy quarks and quarkonium. X. Y.~thanks Nora Brambilla and Jacopo Ghiglieri for useful comments on the manuscript. D.-L. Y.~is supported by National Science and Technology Council (Taiwan) under Grants No. MOST 110-2112-M-001-070-MY3 and No. NSTC 113-2628-M-001-009-MY4. X. Y.~is supported
by the U.S. Department of Energy, Office of Science, Office of Nuclear Physics, InQubator
for Quantum Simulation (IQuS) (https://iqus.uw.edu) under Award No. DOE (NP) DE-SC0020970 via the program on Quantum Horizons: QIS Research and Innovation for Nuclear Science.

\bibliographystyle{jhep}
\bibliography{main}  
\end{document}